\documentclass[12pt,a4paper]{article}
\usepackage[utf8]{inputenc}
\usepackage{amsmath, amssymb, amsthm}
\usepackage{graphicx}
\usepackage{hyperref}
\usepackage{setspace}
\usepackage{subcaption}  
\usepackage{float}      
\usepackage{array}
\usepackage{longtable}
\usepackage{multirow}
\usepackage{amsmath,amssymb,amsfonts} 
\usepackage{cite} 
\usepackage{enumitem} 
\usepackage{caption} 
 \usepackage{float}
\usepackage{booktabs} 
\usepackage{adjustbox}
\usepackage[margin=1in]{geometry} 
\usepackage{authblk}
\usepackage{hyperref}
\title{Beyond the Median Voter Theorem: A New Framework for Ideological Positioning} 
\date{}
\author{Shitong Wang\thanks{\href{shitwang@mail.uni-mannheim.de}{shitwang@mail.uni-mannheim.de}}}
\affil[]{Univeristy of Mannheim}

\begin{document}
\maketitle 

\begin{abstract}
This paper revisits the limitations of the Median Voter Theorem and introduces a novel framework to analyze the optimal economic ideological positions of political parties. By incorporating Nash equilibrium, we examine the mechanisms and elasticity of ideal deviation costs, voter distribution, and policy feasibility. Our findings show that an increase in a party's ideal deviation cost shifts its optimal ideological position closer to its ideal point. Additionally, if a voter distribution can be expressed as a positive linear combination of two other distributions, its equilibrium point must lie within the interval defined by the equilibrium points of the latter two. We also find that decreasing feasibility costs incentivize governments, regardless of political orientation, to increase fiscal expenditures (e.g., welfare) and reduce fiscal revenues (e.g., taxes). This dynamic highlights the fiscal pressures commonly faced by democratic nations under globalization. Moreover, we demonstrate that even with uncertain voter distributions, parties can identify optimal ideological positions to maximize their utility. Lastly, we explain why the proposed framework cannot be applied to community ideologies due to their fundamentally different nature. This study provides new theoretical insights into political strategies and establishes a foundation for future empirical research.


\end{abstract}
\newpage
\section{Introduction}
What economic ideological position should modern political parties adopt in their activities? Since the pioneering works of Hotelling (1929) and Downs (1957), political scientists and economists have widely used spatial models to analyze this question. Hotelling and Downs argued that two competing parties tend to align their ideological positions with the preferences of the median voter, a principle now famously known as the \textit{Median Voter Theorem}. Building on the foundation of this spatial model and Median Voter Theorem, researchers have conducted extensive theoretical and empirical studies to refine and expand its implications.

Palfrey (1984) suggested that third-party candidates could emerge to occupy ideological niches vacated by the two main candidates, depending on their ideological alignment. Enelow and Hinich (1984) highlighted the importance of central ideology in shaping voter behavior within spatial models. Gomez et al. (2007) demonstrated how adverse weather conditions reduce voter turnout. Fenster (1994) showed that same-day voter registration increases electoral participation. Davis and Hinich (1966) and Hinich and Ordeshook (1970) extended the spatial model to a multi-dimensional ideological space, where candidates compete on more than one axis of policy. Coughlin (1992) and Burden (1997) modeled voting as a stochastic process, emphasizing the role of uncertainty in voter behavior and its implications for electoral outcomes. Banks and Duggan (2005) and McKelvey and Patty (2006) further developed probabilistic voting models, demonstrating that candidates’ ideological positions could converge under certain conditions, leading to greater stability in electoral competition and alignment with social welfare maximization. Jones et al. (2022) utilized probabilistic voting rules to develop a model that explains how candidates adjust their positions in response to potential influencing mechanisms. Additionally, Kollman, Miller, and Page (1992) highlighted that candidates are not always solely motivated by winning elections; they may derive utility from securing victory with specific ideological positions.

Furthermore, many scholars have shown that political candidates not only respond to voter preferences but also significantly influence them, especially in polarized environments. This influence extends to voters' ideological views (Druckman et al., 2013) and their affective perceptions of opposing parties (Banda and Cluverius, 2018) and political elites (Rogowski and Langella, 2015). However, Mullinix (2016) emphasized that while top-down social influence is critical, voters do not unconditionally follow polarized elites. This reinforces the relevance of the Median Voter Theorem, as candidates adjusting their ideological positions to align with voter preferences remains a dominant strategy.

In the real world, the ideological divide between political parties has become increasingly pronounced. For instance, in the United Kingdom, political parties are deeply divided, while in the United States, the ideological differences between the Democratic and Republican parties have intensified in recent years. Webster and Abramowitz (2017) proved it through empirical work. This growing polarization manifests not only during elections but also in day-to-day political discourse and governance. As a result, the Median Voter Theorem struggles to account for these contemporary political phenomena. To address this gap, political scientists and economists have conducted extensive theoretical and empirical research.

Black (1958) underscores the importance of single-peaked voter preferences in ensuring the validity of the Median Voter Theorem, but increasing polarization challenges this assumption, necessitating a reevaluation of classical models to better reflect contemporary political dynamics. Fiorina and Abrams (2008) build on this foundation by examining political polarization, demonstrating its role in deepening partisan divisions and shaping voter behavior. Similarly, Baldassarri and Bearman (2007) explore the structural and network mechanisms that drive ideological segmentation, highlighting the sociological underpinnings of polarization. Dixit and Weibull (2007) extend this analysis from an economic perspective, identifying the impact of polarization on collective decision-making and resource allocation. Grosser and Palfrey (2014) take a theoretical approach to explain polarization, developing a model that shows how strategic candidate entry fosters polarization by favoring extreme candidates over moderates. 

Shifting to the role of digital platforms, Conover et al. (2011), Bail et al. (2018), and Wang et al. (2020) investigate how social media amplifies political polarization. Conover et al. (2011) reveal partisan segregation in social networks, while Bail et al. (2018) demonstrate through experiments that exposure to opposing views can intensify polarization rather than reduce it. Wang et al. (2020) complement this work by modeling how echo chambers form through socio-cognitive biases and network effects.

Together, these studies highlight the multifaceted drivers of political polarization, from foundational assumptions about voter behavior to the impact of modern digital and institutional dynamics.

In this paper, we aim to address the limitations of the Median Voter Theorem in explaining the increasing political polarization observed in contemporary settings and proposes a novel theoretical model. We first construct an economic ideology model that incorporates the trade-off between individual voters' motivation to vote and the associated voting costs into the traditional distribution of voter ideologies, thereby clarifying the strategic adjustments political parties make regarding their economic ideological positions. Through this innovation, we introduce the concept of ``ideological deviation costs," capturing the risks faced by parties when deviating from their original ideological stance, including internal fragmentation and the potential loss of their political niche. Furthermore, we examine the theoretical optimal ideological positioning of parties under conditions where the true distribution of voter preferences is unknown. The model is then extended from a one-dimensional framework to a multi-dimensional space by introducing the concept of internal policy feasibility costs, 

Methodologically, we introduce Nash equilibrium into the framework, leveraging theoretical derivations and numerical simulations to systematically analyze the equilibrium properties and the elasticity of party strategies under various influences. This approach ensures the robustness of our theoretical conclusions and provides a comprehensive understanding of the dynamics shaping political polarization. By exploring the equilibrium conditions, we demonstrate how parties respond strategically to changes in voter distributions, ideological deviation costs, and other critical factors. Additionally, the integration of numerical simulations allows us to visualize and quantify these effects, bridging the gap between theoretical insights and empirical observations.

Finally, we explore the significant differences between economic ideology and community ideology, discussing why the latter cannot be directly applied within the theoretical framework of this paper. By offering a new perspective on political polarization and a practical analytical framework, this research provides valuable insights for both theoretical exploration and empirical application.

\section{Theory Preperation and Model Construction}
We begin with an idealized one-dimensional model of economic ideology to provide a foundation for subsequent analysis. 

Let \( \mathcal{P} \) represent the set of candidate parties and \( \mathcal{I} \subseteq \mathbb{R} \) represent the set of economic ideologies associated with these parties. Each party’s economic ideology is assumed to be self-determined, reflecting its inherent values and policy objectives. Define a mapping function \( f \colon  \mathcal{P} \to  \mathcal{I} \) that assigns each candidate party in \( \mathcal{P}\) its respective economic ideology in \( \mathcal{I} \).

Assume that voters’ one-dimensional economic ideologies follow an exogenous, continuous and differentiable distribution \( G(x) \), with support \( \mathbf{X} \subseteq \mathbb{R} \). We further define a support willingness function \( s \colon \mathbf{X} \to \mathcal{P} \), which maps a voter’s ideological position \( x \in \mathbf{X} \) to the candidate party in \( \mathcal{P} \) whose economic ideology is closest to \( x \) in terms of distance. Formally, we define:

\begin{equation}
s(x) = \arg\min_{p \in \mathcal{P}} d(x, f(p)),
\end{equation}
where \( d(x, f(p)) \) is the distance between \( x \) and the economic ideology of candidate party \( p \). In most cases, this distance is taken to be Euclidean. It should be noted that $s(x)$ here does not only refer to voting in a certain election, but can generally refer to support for a certain political party in political activities, such as expressing opinions on social media, participating in demonstrations, etc.

A distinctive feature of this model is that the utility derived from voting is decoupled from the voter’s actual choice of candidate. This partially addresses the \textit{Voter Paradox} which asks why individuals vote when the probability of influencing the election outcome is negligible, and the cost of voting seemingly outweighs any personal benefit. For an individual voter, the distribution of other voters’ ideologies is irrelevant; the critical decision is whether to vote or abstain and, if voting, which candidate to support. 

In several works mentioned above (Coughlin, 1992; Burden, 1997; Banks and Duggan, 2005; McKelvey and Patty, 2006; Jones et al., 2022), authors addressed the uncertainty in voter choice through probabilistic voting models. For example, a left-leaning voter might be modeled as having an 80\% likelihood of voting for a left-wing party and a 20\% likelihood of voting for a right-wing party, reflecting the inherent unpredictability of voting behavior. However, probabilistic voting model may not be suitable for our framework. In our model, we assume that a voter’s economic ideology is the primary determinant of their voting behavior. At the group level, this is also consistent with \textit{Rational Choice Theory}.

This choice pattern implicitly adopts the political science perspective that ideology serves as a heuristic tool, enabling complex policy propositions to be communicated to voters more effectively. Instead of scrutinizing specific policies in detail, voters can grasp the general policy directions based on the economic ideology of parties.

Once voters have identified their preferred party, their subsequent decision depends on whether they are motivated to vote and whether the associated voting cost is acceptable. Voters possess a motivation for voting, denoted by \( m(x) \), which arises from social responsibility, political dignity and desire to express political ideas. The function \( m(x) \) is convex, with its minimum point near the median of the distribution, suggesting that voters with more extreme ideologies are more motivated to vote, while those with moderate ideological positions tend to have lower voting motivation. Voting entails cost \( c \), which represents the opportunity cost of voting, including time and effort. This cost is a random variable independent of the distribution of voters' ideologies and often higher where remote voting is not available.  

Let \( q \colon \mathbf{X} \to \mathcal{P} \cup \{\text{0}\} \) represent a function that maps a voter’s support for a candidate party, considering their voting motivation and cost, to the final voting outcome. Here, \( s(x) \in \mathcal{P} \) symbolizes the candidate a voter supports, and it interacts multiplicatively with \( H(m(x) - c) \), producing the final decision:

\begin{equation}
q(x) = H(m(x) - c) \cdot s(x),
\end{equation}

where \( H(x) \) is the Heaviside function:

\[
H(x) = 
\begin{cases} 
0, & \text{if } x < 0, \\ 
1, & \text{if } x \geq 0.
\end{cases}
\]

Building on this framework, we describe the relationship between a political party’s ideology and the support it receives. Consider a two-party system with a left-wing party (\( \mathit{left} \)) and a right-wing party (\( \mathit{right} \)), such that \( \mathcal{C} = \{ \mathit{left}, \mathit{right} \} \). 

The utility of the left-wing party, denoted as \( U_{\text{left}} \), depends on the ideologies of both the left-wing and right-wing parties, \( f(\mathit{left}) \) and \( f(\mathit{right}) \), respectively. It is expressed as:

\begin{equation}
U_{\text{left}}(f(\mathit{left}), f(\mathit{right})) = E \left( \int_{\mathbf{X}} I_{\text{left}}(q(x)) \, dG(x) \right),
\end{equation}

where \( I_{\text{left}} \) is an indicator function:

\[
I_{\text{left}}(p) = 
\begin{cases} 
1, & \text{if } p = \mathit{left}, \\ 
0, & \text{otherwise}.
\end{cases}
\]

Similarly, the utility function of the right-wing party, \( U_{\text{right}} \), is given by:

\begin{equation}
U_{\text{right}}(f(\mathit{left}), f(\mathit{right})) = E \left( \int_{\mathbf{X}} I_{\text{right}}(q(x)) \, dG(x) \right),
\end{equation}

where \( I_{\text{right}} \) is an indicator function:

\[
I_{\text{right}}(p) = 
\begin{cases} 
1, & \text{if } p = \mathit{right}, \\ 
0, & \text{otherwise}.
\end{cases}
\]

The model described above bears resemblance to the \textit{Hotelling Model}, albeit with certain modifications. According to Median Voter Theorem, in an idealized scenario, both parties would position their ideologies infinitely close to the median of the voter ideological distribution. However, this situation is rarely observed in practice. Downs (1957) attributes this deviation to the risk of alienating extremist supporters. Specifically, if one party moves too close to the other, its extremist base may withdraw support and abstain from voting.

When applied to economic ideology, however, this explanation appears less convincing. Economic ideologies and associated policies tend to have clearer interest alignments and greater adaptability, making compromise more feasible across a spectrum ranging from social democracy to laissez-faire capitalism. For example, a socialist in Great Britain---where socialists here refers to adherents of the Second International who advocate achieving socialism through democratic means such as parliamentary struggle, rather than Leninists who support violent revolution---would unhesitatingly vote for the Labour Party over the Conservative Party. Similarly, in United States, voters who align with the Austrian School of economics would almost certainly vote for the Republican Party, which champions low taxes and small government, over the Democratic Party.

Here we propose an alternative explanation. While voters may not withdraw their support to a party due its ideology moderation, the party itself incurs significant costs. In addition to the risk of another party rising to occupy its original position, the possibility of internal splits is a primary concern. History provides ample evidence of such occurrences.

In the early 20th century, the Democratic Party in the United States experienced significant ideological fragmentation, first between progressives and conservatives. Prominent progressives like Woodrow Wilson and William Jennings Bryan called for increased government intervention in the economy to address issues such as inequality and labor rights, while conservatives within the party championed laissez-faire economic policies. This ideological rift resurfaced in the 1960s, particularly during the Civil Rights Movement, when Southern Democrats, or ``Dixiecrats," opposed federal intervention, culminating in a political realignment that saw many Southern conservatives shift to the Republican Party.

Similarly, in the 1980s, internal divisions within the British Labour Party emerged over economic policy, with the traditional left advocating for socialist policies, including nationalization and strong trade union influence, while the moderates, led by the ``Gang of Four", broke away to form the Social Democratic Party (SDP). The Labour Party’s transformation under Tony Blair in the 1990s, branded as ``New Labour", embraced a centrist stance that reconciled market-driven economics with social justice. This shift alienated some left-wing members, leading to further fragmentation as radical factions formed alternative socialist groups. 

Both cases above exemplify how shifts in ideological orientation can precipitate significant party realignments and divisions. We can conclude that when a political party deviates too far from its original ideology, it risks internal division, which may cause a whole utility reduction. Therefore, for a political party, choosing an economic ideology that maximizes support is of great importance.

To capture this, we modify the utility functions of the two parties as follows:
\begin{equation}
U_{\text{left}}(f(\mathit{left}), f(\mathit{right})) = E \left( \int_{\mathbf{X}} I_{\text{left}}(q(x)) \, dG(x) \right) - D_\text{left}(f(\mathit{left}), \mathit{ideal}_{\text{left}}), 
\end{equation}
\begin{equation}
U_{\text{right}}(f(\mathit{left}), f(\mathit{right})) = E \left( \int_{\mathbf{X}} I_{\text{right}}(q(x)) \, dG(x) \right) - D_\text{right}(f(\mathit{right}), \mathit{ideal}_{\text{right}}), 
\end{equation}
where \( \mathit{ideal}_{\text{left}} \) and \( \mathit{ideal}_{\text{right}} \) represent the original economic ideologies of the \( \mathit{left} \) and \( \mathit{right} \) parties, respectively. Here, \( D_{\text{left}}(f(\mathit{left}), \mathit{ideal}_{\text{left}}) \) and \( D_{\text{right}}(f(\mathit{right}), \mathit{ideal}_{\text{right}}) \) are functions that measure the cost of deviating from each party's original ideological position. 

Suppose \( D_{\text{left}}(f(\mathit{left}), \mathit{ideal}_{\text{left}}) \) and \( D_{\text{right}}(f(\mathit{right}), \mathit{ideal}_{\text{right}}) \) are convex and differentiable functions, defined for fixed values of \( \mathit{ideal}_{\text{left}} \) and \( \mathit{ideal}_{\text{right}} \). Each function achieves its minimum precisely at \( \mathit{ideal}_{\text{left}} \) or \( \mathit{ideal}_{\text{right}} \), respectively.

\section{One Dimension Model Properties Analysis}
In this section, we examine the properties of the model presented above and investigate the internal political mechanisms underpinning its structure and dynamics. Without loss of generality, we define the strategy space as $\mathbf{X} = [-1, 1]$. For simplicity, we abbreviate $f(\mathit{left})$ and $f(\mathit{right})$ as $x_{\text{left}}$ and $x_{\text{right}}$, respectively. Accordingly, the utility functions of the two parties are expressed as:

\[
U_{\mathit{left}}(x_{\text{left}}, x_{\text{right}}) = \int_{-1}^{\frac{x_{\text{left}} + x_{\text{right}}}{2}} g(x) F_c(m(x)) \, dx - D_{\mathit{left}}(x_{\text{left}}, \mathit{ideal}_{\mathit{left}}),\tag{4}
\]
\[
U_{\mathit{right}}(x_{\text{left}}, x_{\text{right}}) = \int_{\frac{x_{\text{left}} + x_{\text{right}}}{2}}^{1} g(x) F_c(m(x)) \, dx - D_{\mathit{right}}(x_{\text{right}}, \mathit{ideal}_{\mathit{right}}),\tag{5}
\]
where \( g(x) \) is the probability density function (PDF) of the distribution \( G \), and \( F_c(x) \) is the cumulative distribution function (CDF) of the random variable \( c \).

To facilitate further analysis, we derive the partial derivatives of the utility functions 
with respect to $x_{\text{left}}$ and $x_{\text{right}}$, which are expressed as follows:
\begin{equation}
\frac{\partial U_{\text{left}}}{\partial x_{\text{left}}} = \frac{1}{2} g\left( \frac{x_{\text{left}} + x_{\text{right}}}{2} \right) F_c\left( m\left( \frac{x_{\text{left}} + x_{\text{right}}}{2} \right) \right) - \frac{\text{d} D_{\text{left}}}{\text{d} x_{\text{left}}}(x_{\text{left}}, \mathit{ideal}_{\text{left}}),
\end{equation}
\begin{equation}
\frac{\partial U_{\text{left}}}{\partial x_{\text{right}}} = \frac{1}{2} g\left( \frac{x_{\text{left}} + x_{\text{right}}}{2} \right) F_c\left( m\left( \frac{x_{\text{left}} + x_{\text{right}}}{2} \right) \right),
\end{equation}
\begin{equation}
\frac{\partial U_{\text{right}}}{\partial x_{\text{right}}} = -\frac{1}{2} g\left( \frac{x_{\text{left}} + x_{\text{right}}}{2} \right) F_c\left( m\left( \frac{x_{\text{left}} + x_{\text{right}}}{2} \right) \right) - \frac{\text{d} D_{\text{right}}}{\text{d} x_{\text{right}}}(x_{\text{right}}, \mathit{ideal}_{\text{right}}),
\end{equation}
\begin{equation}
\frac{\partial U_{\text{right}}}{\partial x_{\text{left}}} = -\frac{1}{2} g\left( \frac{x_{\text{left}} + x_{\text{right}}}{2} \right) F_c\left( m\left( \frac{x_{\text{left}} + x_{\text{right}}}{2} \right) \right).
\end{equation}

\subsection{Best Response Variation}

First, we analyze how one party’s best response adjusts in relation to the other party’s strategy. 
Let $U_{\text{left}}$ and $U_{\text{right}}$ be continuous and quasi-concave utility functions. 
To illustrate, we consider the variation of $x_{\text{left}}$ with respect to $x_{\text{right}}$. 
When $x_{\text{right}}$ changes, implicit differentiation is applied to determine how $x_{\text{left}}$ 
should adjust to maximize $U_{\text{left}}$:

\begin{equation}
\frac{dx_{\text{right}}}{dx_{\text{left}}} = 
-\frac{\frac{\partial^2 U_\text{left}}{\partial x_\text{left}^2}}
{\frac{\partial^2 U_\text{left}}{\partial x_\text{left} \partial x_\text{right}}} 
= 
-\frac{\frac{\partial}{\partial x_{\text{left}}} 
\left( \frac{1}{2} g\left( \frac{x_{\text{left}} + x_{\text{right}}}{2} \right) 
F_c\left( m\left( \frac{x_{\text{left}} + x_{\text{right}}}{2} \right) \right) 
- \frac{\partial D_{\text{left}}}{\partial x_{\text{left}}} \right)}
{\frac{\partial}{\partial x_{\text{right}}} 
\left( \frac{1}{2} g\left( \frac{x_{\text{left}} + x_{\text{right}}}{2} \right) 
F_c\left( m\left( \frac{x_{\text{left}} + x_{\text{right}}}{2} \right) \right) 
- \frac{\partial D_{\text{left}}}{\partial x_{\text{left}}} \right)}.
\end{equation}

With $\frac{\partial^2 U_\text{left}}{\partial x_\text{left}^2}<0$ at the maximum point, $\frac{dx_{\text{right}}}{dx_{\text{left}}} > 0$ if and only if:
\begin{equation}
\frac{\partial}{\partial x_{\text{left}}} 
\left( \frac{1}{2} g\left( \frac{x_{\text{left}} + x_{\text{right}}}{2} \right) 
F_c\left( m\left( \frac{x_{\text{left}} + x_{\text{right}}}{2} \right) \right) \right) > 0.
\end{equation}
Conversely, $\frac{dx_{\text{right}}}{dx_{\text{left}}} < 0$ if and only if: 
\begin{equation}
\frac{\partial}{\partial x_{\text{left}}} 
\left( \frac{1}{2} g\left( \frac{x_{\text{left}} + x_{\text{right}}}{2} \right) 
F_c\left( m\left( \frac{x_{\text{left}} + x_{\text{right}}}{2} \right) \right) \right) < 0.
\end{equation}
The variation of $x_{\text{left}}$ with respect to $x_{\text{right}}$ can be derived analogously.

\subsection{Existence and Uniqueness of Nash Equilibrium}

The structure of the utility functions naturally raises a fundamental question in game theory and political science: the existence and uniqueness of Nash equilibrium in the ideological positions of two parties.

According to \textit{Glicksberg’s Theorem}, at least one Nash equilibrium exists under the following conditions:

\begin{enumerate}
    \item \textbf{Non-empty, Compact, and Convex Strategy Spaces}: Each player’s strategy space must be a non-empty, compact, and convex subset of a locally convex topological vector space.
    \item \textbf{Continuity of Utility Functions}: Each player’s utility function must be continuous in the joint strategy profile.
\end{enumerate}

Let us examine these conditions in the context of the model.
\subsection*{Condition 1: Strategy Space}

The strategy set \( \mathbf{X} = [-1, 1] \) is non-empty, compact, and convex in the topological space of \( \mathbb{R} \). Moreover, any linear combination of two strategies lies within the strategy space. Thus, \textbf{Condition 1} is satisfied.

\subsection*{Condition 2: Continuity of Utility Functions}

The continuity of the utility functions is guaranteed by the continuity of \( g(x) \), \( F_c(m(x)) \), and \( D_{\text{left}}, D_{\text{right}} \). Therefore, \textbf{Condition 2} is also satisfied.

Hence, there exists at least one pair of Nash equilibrium points \( x_{\text{left}}^* \) and \( x_{\text{right}}^* \) in our model such that:
\begin{equation}
U_{\text{left}}(x_{\text{left}}^*, x_{\text{right}}^*) \geq U_{\text{left}}(x_{\text{left}}, x_{\text{right}}^*),
\end{equation}
\begin{equation}
U_{\text{right}}(x_{\text{left}}^*, x_{\text{right}}^*) \geq U_{\text{right}}(x_{\text{left}}^*, x_{\text{right}}),
\end{equation}
for all \( x_{\text{left}} \) and \( x_{\text{right}} \) in the strategy space.

\subsection*{Condition 3: Quasi-concavity of Utility Functions}

To ensure the uniqueness of Nash equilibrium, an additional condition is required: strict quasi-concavity of utility functions. Strict quasi-concavity guarantees that each player’s best response is uniquely determined for any fixed strategy of the opposing player. Consequently, the intersection of the players’ best response functions results in a unique pair of Nash equilibrium points.

Moreover, when the equilibrium point lies in the interior of the strategy set, the first derivative of the utility function with respect to the player’s strategy is zero, and the second derivative is strictly negative at this point, satisfying the local maximum condition.

Under the assumption that the conditions for the existence and uniqueness of Nash equilibrium are satisfied, we now proceed to analyze the properties of this unique equilibrium.

\subsection{Basic Equilibrium Distribution}
First, we analyze the basic equilibrium distribution of the Nash equilibrium. Let \( (x_{\text{left}}^*, x_{\text{right}}^*) \) denote the unique Nash equilibrium when the original economic ideologies of the two parties are \( \mathit{ideal}_\text{left} \) and \( \mathit{ideal}_\text{right}\), respectively. Then, the relationship between these points is given by:
\begin{equation}
\mathit{ideal}_\text{left} \leq x_{\text{left}}^* \leq x_{\text{right}}^* \leq \mathit{ideal}_\text{right}.
\end{equation}

When \( x_{\text{left}} \leq \mathit{ideal}_\text{left} \), we have \( \frac{\partial U_{\text{left}}}{\partial x_{\text{left}}} \geq 0 \), indicating that any \( x_{\text{left}} \leq ideal_\text{left} \) is not the best response to a given \( x_{\text{right}} \). Similarly, when \( x_\text{right} \geq \mathit{ideal}_{\mathit{right}} \), we have \( \frac{\partial U_{\text{right}}}{\partial x_\text{right}} \leq 0 \), implying that any \( x_{\text{right}} \geq \mathit{ideal}_\text{right} \) is not the best response to a given \( x_{\text{left}} \). 

This conclusion aligns with real-world political dynamics: while political parties may advocate for idealistic ideologies, they are compelled to adopt more moderate positions in practice to garner broader electoral support. Additionally, when the deviation cost functions of the two parties are symmetrical about the \( y \)-axis, the resulting Nash equilibrium points are also symmetrical about the \( y \)-axis. In this case, the equilibrium point values are determined by the term \( g(0)F_c(m(0)) \). The figures below illustrate the distribution of Nash equilibrium points when the deviation cost functions are symmetrical about the \( y \)-axis. We consider three representative voter distributions with symmetrical deviation cost functions: left-skewed, normal, and right-skewed. Evidently, the resulting pairs of Nash equilibria are symmetrical about the \( y \)-axis.

\begin{figure}[H]
    \centering
    \begin{minipage}[b]{0.28\textwidth}
        \centering
        \includegraphics[width=\textwidth]{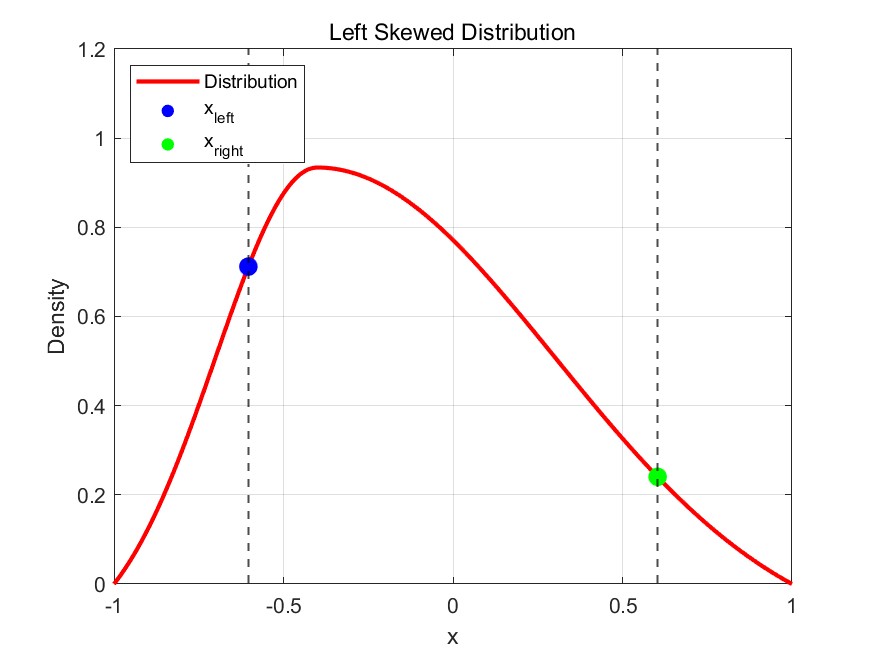}
        \caption{Nash Equilibrium with Left Skewed Distribution}
        \label{fig:image1}
    \end{minipage}
    \hspace{0.05\textwidth}  
    \begin{minipage}[b]{0.28\textwidth}
        \centering
        \includegraphics[width=\textwidth]{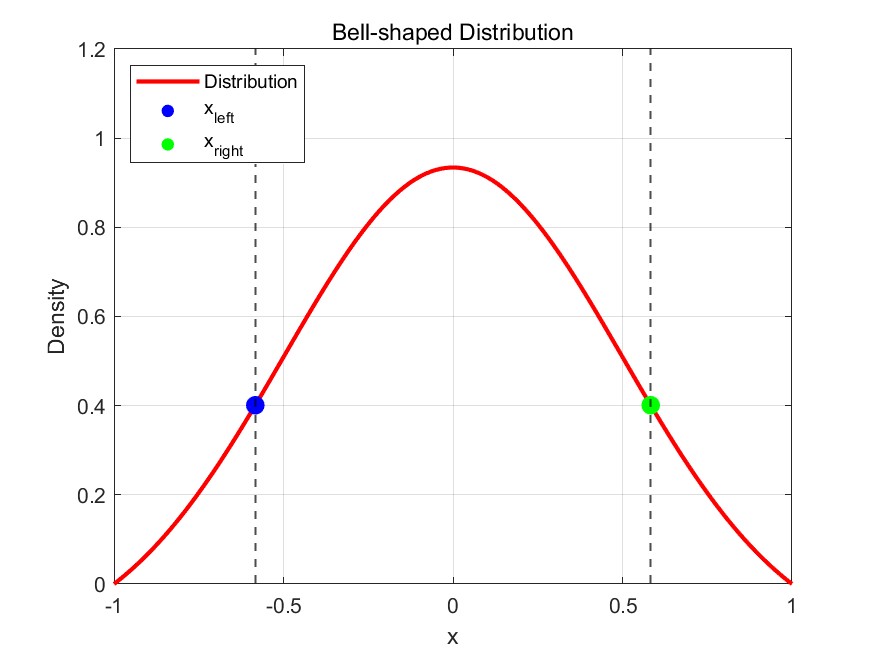}
        \caption{Nash Equilibrium with Bell-shaped Distribution }
        \label{fig:image2}
    \end{minipage}
    \hspace{0.05\textwidth}  
    \begin{minipage}[b]{0.28\textwidth}
        \centering
        \includegraphics[width=\textwidth]{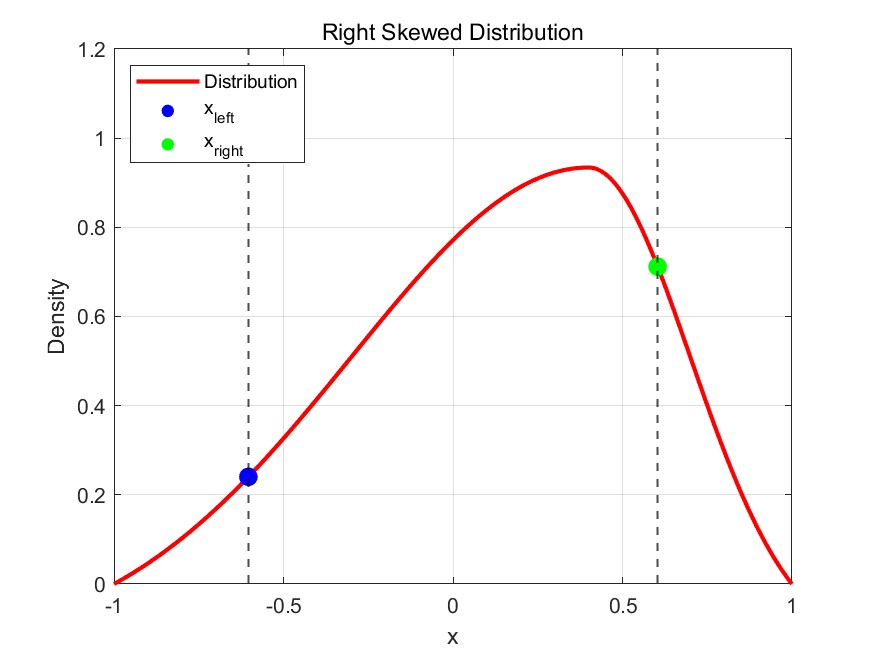}
        \caption{Nash Equilibrium with Right Skewed Distribution}
        \label{fig:image3}
    \end{minipage}
\end{figure}

\subsection{Deviation Cost Elasticity of Equilibrium}

In this section, we analyze the elasticity of equilibrium positions with respect to changes in deviation costs. As previously established, the equilibrium points \( x_{\text{left}}\) and \( x_{\text{right}} \) satisfy the following conditions:
\begin{equation}
\frac{\partial U_{\text{left}}}{\partial x_{\text{left}}}(x_{\text{left}}^*, x_{\text{right}}^*) = 0,
\end{equation}
\begin{equation}
\frac{\partial U_{\text{right}}}{\partial x_{\text{right}}}(x_{\text{left}}^*, x_{\text{right}}^*) = 0,
\end{equation}
\begin{equation}
\frac{\partial U_{\text{left}}^2}{\partial^2 x_{\text{left}}}(x_{\text{left}}^*, x_{\text{right}}^*) < 0,
\end{equation}
\begin{equation}
\frac{\partial U_{\text{right}}^2}{\partial^2 x_{\text{right}}}(x_{\text{left}}^*, x_{\text{right}}^*) < 0.
\end{equation}
 
Next, we introduce a small perturbation \(\epsilon > 0\) to the deviation cost function \( D_\text{right} \). This perturbation alters the equilibrium conditions, causing shifts in the equilibrium points \( x_{\text{left}}^*\) and \( x_{\text{right}}^* \) by \(\Delta x_{\text{left}}^*\) and \(\Delta x_{\text{right}}^*\), respectively. To quantify these shifts, we employ a first-order Taylor expansion around the original equilibrium points \( x_{\text{left}}^*\) and \( x_{\text{right}}^*\). The perturbed partial derivatives of the utility functions can then be expressed as:
\begin{equation}
\frac{\partial U_{\text{left}}}{\partial x_{\text{left}}} \approx \frac{\partial^2 U_{\text{left}}}{\partial x_{\text{left}}^2} \Delta x_{\text{left}} + \frac{\partial^2 U_{\text{left}}}{\partial x_{\text{left}} \partial x_{\text{right}}} \Delta x_{\text{right}},
\end{equation}
\begin{equation}
\frac{\partial U_{\text{right}}}{\partial x_{\text{right}}} \approx \frac{\partial^2 U_{\text{right}}}{\partial x_{\text{right}}^2} \Delta x_{\text{right}} + \frac{\partial^2 U_{\text{right}}}{\partial x_{\text{right}} \partial x_{\text{left}}} \Delta x_{\text{left}}.
\end{equation}

Substituting these into the perturbed equilibrium conditions, we form the following linear system:
\begin{equation}
\frac{\partial^2 U_{\text{left}}}{\partial x_{\text{left}}^2} \Delta x_{\text{left}} + \frac{\partial^2 U_{\text{left}}}{\partial x_{\text{left}} \partial x_{\text{right}}} \Delta x_{\text{right}} = 0, 
\end{equation}
\begin{equation}
\frac{\partial^2 U_{\text{right}}}{\partial x_{\text{right}} \partial x_{\text{left}}} \Delta x_{\text{left}}+ \frac{\partial^2 U_{\text{right}}}{\partial x_{\text{right}}^2} \Delta x_{\text{right}} = \epsilon \frac{\partial D_{\text{right}}}{\partial x_{\text{right}}}.
\end{equation}

This system can be written in matrix form:
\begin{equation}
\mathbf{H} \cdot \Delta \mathbf{x}  = \mathbf{b} ,
\end{equation}
where
\[
\mathbf{H} =
\begin{bmatrix}
\frac{\partial^2 U_{\text{left}}}{\partial x_{\text{left}}^2} & \frac{\partial^2 U_{\text{left}}}{\partial x_{\text{left}} \partial x_{\text{right}}} \\
\frac{\partial^2 U_{\text{right}}}{\partial x_{\text{right}} \partial x_{\text{left}}} & \frac{\partial^2 U_{\text{right}}}{\partial x_{\text{right}}^2}
\end{bmatrix}, \Delta \mathbf{x}=\begin{bmatrix} \Delta x_{\text{left}} \\ \Delta x_{\text{right}} \end{bmatrix}, \mathbf{b}=\begin{bmatrix} 0 \\ \epsilon \frac{\partial D_{\text{right}}}{\partial x_{\text{right}}} \end{bmatrix}.\]

Solving the linear system gives:
\[
\begin{bmatrix} \Delta x_{\text{left}} \\ \Delta x_{\text{right}} \end{bmatrix} = \mathbf{H}^{-1} \cdot \begin{bmatrix} 0 \\ \epsilon \frac{\partial D_{\text{right}}}{\partial x_{\text{right}}} \end{bmatrix}.
\]
Then we get
\begin{equation}
\Delta x_{\text{left}} = \frac{-\epsilon H_{12} \cdot \left( \frac{\partial D_{\text{right}}}{\partial x_{\text{right}}} \right)}{\det(\mathbf{H})},
\end{equation}
\begin{equation}
\Delta x_{\text{right}} = \frac{\epsilon H_{11} \cdot \left( \frac{\partial D_{\text{right}}}{\partial x_{\text{right}}} \right)}{\det(\mathbf{H})}.
\end{equation}
det$(\mathbf{H})= \frac{\partial^2 U_{\text{left}}}{\partial x_{\text{left}}^2}\cdot\frac{\partial^2 U_{\text{right}}}{\partial x_{\text{right}}^2}- \frac{\partial^2 U_{\text{left}}}{\partial x_{\text{left}} \partial x_{\text{right}}}\cdot \frac{\partial^2 U_{\text{right}}}{\partial x_{\text{right}} \partial x_{\text{left}}}  >0$ always establishes because $\frac{\partial^2 U_{\text{left}}}{\partial x_{\text{left}}^2}, \frac{\partial^2 U_{\text{right}}}{\partial x_{\text{left}}^2}<0$ while $\frac{\partial^2 U_{\text{right}}}{\partial x_{\text{right}} \partial x_{\text{left}}}=-\frac{\partial^2 U_{\text{left}}}{\partial x_{\text{right}} \partial x_{\text{left}}}$.

Let us abbreviate \(\frac{\partial^2 U_{\text{left}}}{\partial x_{\text{left}}\partial x_{\text{right}}}=-\frac{\partial^2 U_{\text{right}}}{\partial x_{\text{left}}\partial x_{\text{right}}}=\mathcal{L}(x_{\text{left}}, x_{\text{right}}) \), then we can express det$(\mathbf{H})$ as:
\begin{equation}
\begin{aligned}
\det(\mathbf{H}) &= 
\left( \mathcal{L}(x_{\text{left}}, x_{\text{right}}) - \frac{\partial^2 D_{\text{left}}}{\partial x_{\text{left}}^2} \right)
\left( -\mathcal{L}(x_{\text{right}}, x_{\text{right}}) - \frac{\partial^2 D_{\text{right}}}{\partial x_{\text{right}}^2} \right) \\
&\quad - \mathcal{L}(x_{\text{left}}, x_{\text{right}}) \cdot \left( -\mathcal{L}(x_{\text{left}}, x_{\text{right}}) \right) \\
&= \mathcal{L}(x_{\text{left}}, x_{\text{right}}) \cdot 
\left( \frac{\partial^2 D_{\text{left}}}{\partial x_{\text{left}}^2} - \frac{\partial^2 D_{\text{right}}}{\partial x_{\text{right}}^2} \right).
\end{aligned}
\end{equation}

With this, we further simplify the expressions for the shifts in equilibrium points:
\begin{equation}
\Delta x_{\text{left}} = \frac{-\epsilon \frac{\partial D_{\text{right}}}{\partial x_{\text{right}}}}{\frac{\partial^2 D_{\text{left}}}{\partial x_{\text{left}}^2} - \frac{\partial^2 D_{\text{right}}}{\partial x_{\text{right}}^2}} ,\tag{26}
\end{equation}
\begin{equation}
\Delta x_{\text{right}} = \frac{\epsilon \frac{\partial D_{\text{right}}}{\partial x_{\text{right}}} \left( \mathcal{L}(x_{\text{left}}, x_{\text{right}}) - \frac{\partial^2 D_{\text{left}}}{\partial x_{\text{left}}^2} \right)}{\mathcal{L}(x_{\text{left}}, x_{\text{right}}) \cdot 
\left( \frac{\partial^2 D_{\text{left}}}{\partial x_{\text{left}}^2} - \frac{\partial^2 D_{\text{right}}}{\partial x_{\text{right}}^2} \right)}.\tag{27}
\end{equation}

To conclude, we get \(\Delta x_{\text{left}} > 0\) if and only if $\frac{\partial^2 D_{\text{left}}}{\partial x_{\text{left}}^2} > \frac{\partial^2 D_{\text{right}}}{\partial x_{\text{right}}^2};$
\(\Delta x_{\text{left}} < 0\) if and only if $ \frac{\partial^2 D_{\text{left}}}{\partial x_{\text{left}}^2} < \frac{\partial^2 D_{\text{right}}}{\partial x_{\text{right}}^2}; $
\(\Delta x_{\text{right}} > 0\) always establishes for det$(\mathbf{H})>0$, $\frac{\partial^2 U_{\text{left}}}{\partial x_{\text{left}}^2}<0$ and $\frac{\partial D_{\text{right}}}{\partial x_{\text{right}}}<0$. Therefore, when the cost of deviation for one party increases, its ideological position will always shift closer to its ideal point.

Thus, the elasticity of equilibrium with respect to the deviation cost perturbation can be expressed as:
\begin{equation}
E_{x_{\text{left}}} = \frac{- \frac{\partial D_{\text{right}}}{\partial x_{\text{right}}}}{\frac{\partial^2 D_{\text{left}}}{\partial x_{\text{left}}^2} - \frac{\partial^2 D_{\text{right}}}{\partial x_{\text{right}}^2}}\cdot \frac{1}{x_{\text{left}}^*},
\end{equation}
\begin{equation}
E_{x_{\text{right}}} = \frac{ \frac{\partial D_{\text{right}}}{\partial x_{\text{right}}} \left( \mathcal{L}(x_{\text{left}}, x_{\text{right}}) - \frac{\partial^2 D_{\text{left}}}{\partial x_{\text{left}}^2} \right)}{\mathcal{L}(x_{\text{left}}, x_{\text{right}}) \cdot 
\left( \frac{\partial^2 D_{\text{left}}}{\partial x_{\text{left}}^2} - \frac{\partial^2 D_{\text{right}}}{\partial x_{\text{right}}^2} \right)}\cdot \frac{1}{x_{\text{right}}^*}. 
\end{equation}

In a similar manner, elasticity of equilibrium with respect to perturbations in \(D_{\text{left}}\) can be derived through the same methodology.

To further verify the perturbation of deviation costs on the Nash equilibrium, we take a numerical example. Assume the voter distribution follows a common bell-shaped distribution: $g(x) = \frac{1}{0.9256}(e^{-\frac{x^2}{2 \cdot 0.5^2}} - e^{-2})$, where the support of the distribution is \( [-1, 1] \). Suppose \( Fc(x) = 0.5x \) and \( m(x) = x^2 + 1 \), with \( D_{\text{left}} = k_{\text{left}}(x_{\text{left}} + 0.7)^2 \) and \( D_{\text{right}} = k_{\text{right}}(x_{\text{right}} - 0.7)^2 \). Setting \( k_{\text{left}} = 0.6 \), we analyze the variation in the Nash equilibrium points with respect to \( k_{\text{right}} \).

\begin{figure}[H]
    \centering
    \begin{minipage}[b]{0.4\textwidth}
        \centering
        \includegraphics[width=\textwidth]{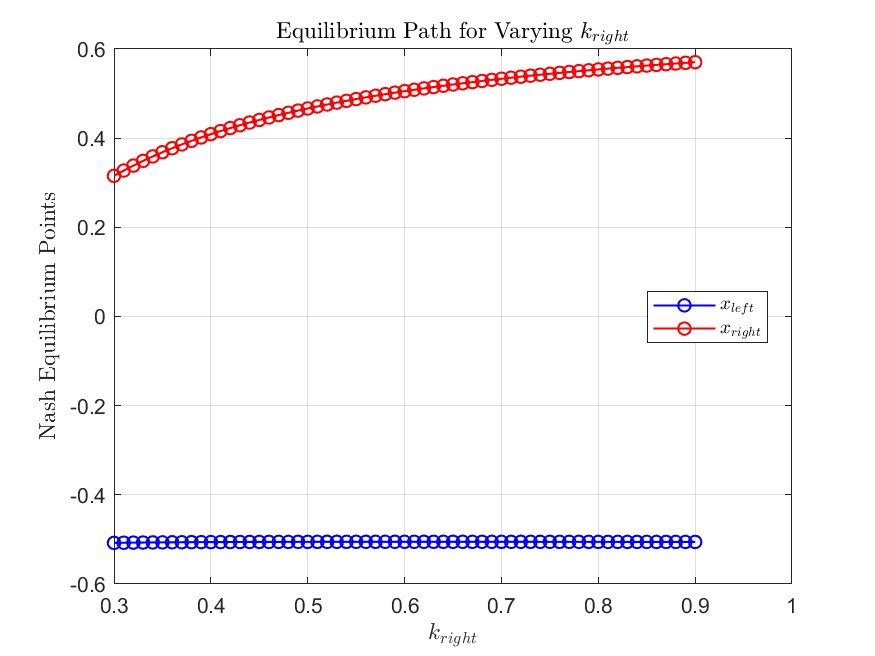}
        \caption{Equilibrium Path for Varying $k_{\text{right}}$}
        \label{fig:image1}
    \end{minipage}
    \hspace{0.05\textwidth}  
    \begin{minipage}[b]{0.4\textwidth}
        \centering
        \includegraphics[width=\textwidth]{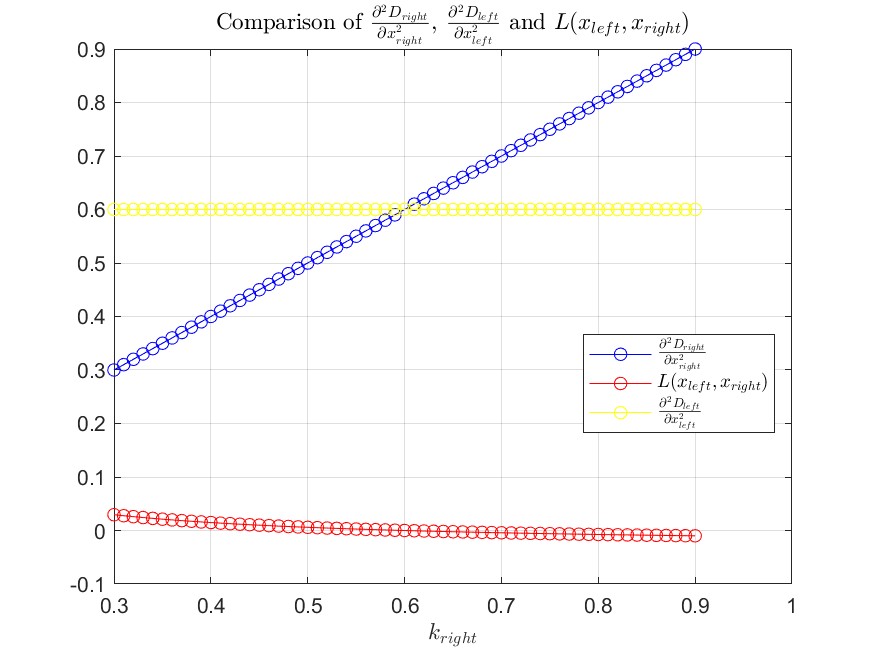}
        \caption{Comparison of Derivatives and $L(x_{\text{left}},x_{\text{right}})$}
        \label{fig:image2}
    \end{minipage}
 
\end{figure}

We observe from the variation of Nash equilibrium points that when \( k_\text{right} < k_\text{left} \), i.e., \( \frac{\partial^2 D_{\text{left}}}{\partial x_{\text{left}}^2} > \frac{\partial^2 D_{\text{right}}}{\partial x_{\text{right}}^2} \), it follows that \( \Delta x_{\text{left}} > 0 \). Conversely, when \( k_\text{right} > k_\text{left} \), i.e., \( \frac{\partial^2 D_{\text{left}}}{\partial x_{\text{left}}^2} < \frac{\partial^2 D_{\text{right}}}{\partial x_{\text{right}}^2} \), we find that \( \Delta x_{\text{left}} < 0 \).

For $\Delta x_\text{right}$, when \( k_\text{right} < k_\text{left} \), i.e., \( \frac{\partial^2 D_{\text{left}}}{\partial x_{\text{left}}^2} > \frac{\partial^2 D_{\text{right}}}{\partial x_{\text{right}}^2} \) and \( \frac{\partial^2 D_{\text{left}}}{\partial x_{\text{left}}^2} > \mathcal{L}(x_{\text{left}}, x_{\text{right}}) > 0 \), then \( \Delta x_{\text{right}} > 0 \). When \( k_\text{right} > k_\text{left} \), i.e., \( \frac{\partial^2 D_{\text{left}}}{\partial x_{\text{left}}^2} < \frac{\partial^2 D_{\text{right}}}{\partial x_{\text{right}}^2} \) and \( \mathcal{L}(x_{\text{left}}, x_{\text{right}}) < 0 \), then \( \Delta x_{\text{right}} > 0 \).   

These phenomena are completely consistent with the condition above.

\subsection{Voter Distribution Elastcity of of Equilibrium}
Now, we examine the elasticity of the Nash equilibrium with respect to the ideological distribution of voters. We introduce a small perturbation \(\gamma > 0\) in the ideological distribution. Let the new distribution \(h(x)\) over the interval \([-1, 1]\) satisfy the condition \(\int_{-1}^{1} h(x) dx = 0\). The perturbed distribution is then given by \(g(x) + \gamma h(x)\). Similar to our previous analysis of small perturbations in deviation costs, we analyze the effect of small perturbations in the ideological distribution of voters.

To simplify notation, we define: $\mathcal{K}(x_\text{left}, x_\text{right}) = \frac{\partial \left(\frac{1}{2} h\left(\frac{x_\text{left} + x_\text{right}}{2}\right) F_c\left(m\left(\frac{x_\text{left} + x_\text{right}}{2}\right)\right)\right)}{\partial x_\text{left}}
= \frac{\partial \left(\frac{1}{2} h\left(\frac{x_\text{left} + x_\text{right}}{2}\right) F_c\left(m\left(\frac{x_\text{left} + x_\text{right}}{2}\right)\right)\right)}{\partial x_\text{right}}.$ Denote the Nash equilibrium points as \(x_\text{left}^g\) and \(x_\text{right}^g\) under the voter distribution \(g(x)\). The perturbation induces shifts in the equilibrium points, resulting in deviations \(\Delta x_\text{left}^g\) and \(\Delta x_\text{right}^g\). Using a first-order Taylor expansion around the original equilibrium, the perturbed partial derivatives of the utility functions can be approximated as:  
\begin{equation}
\frac{\partial^2 U_\text{left}}{\partial x_\text{left}^2} \Delta x_\text{left}^g + \frac{\partial^2 U_\text{left}}{\partial x_\text{left} \partial x_\text{right}} \Delta x_\text{right}^g = -\epsilon \mathcal{K}(x_\text{left}^g, x_\text{right}^g),
\end{equation}
\begin{equation}
\frac{\partial^2 U_\text{right}}{\partial x_\text{right} \partial x_\text{left}} \Delta x_\text{left}^g + \frac{\partial^2 U_\text{right}}{\partial x_\text{right}^2} \Delta x_\text{right}^g = \epsilon \mathcal{K}(x_\text{left}^g, x_\text{right}^g).
\end{equation}

This system can be written in matrix form:
\begin{equation}
\mathbf{H} \cdot \begin{bmatrix} \Delta x_{\text{left}}^g \\ \Delta x_{\text{right}}^g \end{bmatrix} = \begin{bmatrix} -\gamma \mathcal{K}(x_\text{left}^g,x_\text{right}^g) \\ \gamma \mathcal{K}(x_\text{left}^g,x_\text{right}^g) \end{bmatrix},
\end{equation}
then we get
\begin{equation}
\Delta x_{\text{left}}^g = \frac{\gamma \mathcal{K}(x_\text{left}^g,x_\text{right}^g) \left( \frac{d^2 D_{\text{right}}}{d x^2_{\text{right}}} \right)}{\det(\mathbf{H})},
\end{equation}
\begin{equation}
\Delta x_{\text{right}}^g = -\frac{\gamma \mathcal{K}(x_\text{left}^g,x_\text{right}^g) \left( \frac{d^2 D_{\text{left}}}{d x^2_{\text{left}}} \right)}{\det(\mathbf{H})}.
\end{equation}

Thus, the elasticity of equilibrium with respect to the voter distribution perturbation can be expressed as:
\begin{equation}
E_{x_{\text{left}}^g} = \frac{  \mathcal{K}(x_\text{left}^g,x_\text{right}^g) \left( \frac{d^2 D_{\text{right}}}{d x^2_{\text{right}}} \right)}{\det(\mathbf{H})}\cdot \frac{1}{x_{\text{left}}^g},
\end{equation}
\begin{equation}
E_{x_{\text{right}}^g} = -\frac{  \mathcal{K}(x_\text{left}^g,x_\text{right}^g) \left( \frac{d^2 D_{\text{left}}}{d x^2_{\text{left}}} \right)}{\det(\mathbf{H})}\cdot \frac{1}{x_{\text{right}}^g}.
\end{equation}

It can be noted that any perturbation in the voter distribution always leads to equilibrium points shifting towards opposite directions.

Now, we prove that for any three distributions \(g\), \(f\), and \(s\), satisfying \(x_\text{left}^g > x_\text{left}^s > x_\text{left}^f\), it must also hold that \(x_\text{right}^g < x_\text{right}^s < x_\text{right}^f\). Given \(x_\text{left}^g > x_\text{left}^s > x_\text{left}^f\) and the convexity of \(D_\text{left}\), we have:
\begin{equation}
\frac{d D_\text{left}}{d x_\text{left}} \bigg|_{x_\text{left} = x_\text{left}^g} 
> \frac{d D_\text{left}}{d x_\text{left}} \bigg|_{x_\text{left} = x_\text{left}^s} 
> \frac{d D_\text{left}}{d x_\text{left}} \bigg|_{x_\text{left} = x_\text{left}^f}.
\end{equation}
Consequently, by equilibrium conditions:
\begin{equation}
\frac{d D_\text{right}}{d x_\text{right}} \bigg|_{x_\text{right} = x_\text{right}^g} 
< \frac{d D_\text{right}}{d x_\text{right}} \bigg|_{x_\text{right} = x_\text{right}^s} 
< \frac{d D_\text{right}}{d x_\text{right}} \bigg|_{x_\text{right} = x_\text{right}^f}.
\end{equation}
Since \(D_\text{right}\) is also convex, it follows that \(x_\text{right}^g < x_\text{right}^s < x_\text{right}^f\).

Next, consider two voter ideological distributions \(g(x)\) and \(s(x)\), satisfying \(\int_{-1}^{1} g(x) dx = 1\) and \(\int_{-1}^{1} s(x) dx = 1\). A linear combination of these distributions is given by \(h(x) = (1-\lambda) g(x) + \lambda s(x)\), where \(0 < \lambda < 1\), and it also satisfies \(\int_{-1}^{1} h(x) dx = 1\). Suppose \(x_\text{left}^g > x_\text{left}^s\) and \(x_\text{right}^g < x_\text{right}^s\). Then, for the mixed distribution \(h(x)\), it must hold that: $x_\text{left}^g > x_\text{left}^{h} > x_\text{left}^s, \quad x_\text{right}^g < x_\text{right}^{h} < x_\text{right}^s.$
This can be easily proven by contradiction. Assume that \(x_\text{left}^{h}\) or \(x_\text{right}^{h}\) lies outside the intervals \((x_\text{left}^s, x_\text{left}^g)\) or \((x_\text{right}^g, x_\text{right}^s)\). Since the variation of equilibrium points is continuous as \(\lambda\) changes, there must exist some \(\lambda\) such that:
$
x_\text{left}^{h} = x_\text{left}^g$, $x_\text{right}^{h} = x_\text{right}^g$ or $ x_\text{left}^{h} = x_\text{left}^s$, $x_\text{right}^{h} = x_\text{right}^s$.

Taking \(x_\text{left}^{h} = x_\text{left}^g\) and \(x_\text{right}^{h} = x_\text{right}^g\) as an example, we get the following condition:
 
\begin{align}
& \frac{1}{2} ((1-\lambda)g+\lambda s) 
\left( \frac{x_\text{left}^{(1-\lambda)g+\lambda s} + x_\text{right}^{(1-\lambda)g+\lambda s}}{2} \right)
F_c\left( m\left( \frac{x_\text{left}^{(1-\lambda)g+\lambda s} + x_\text{right}^{(1-\lambda)g+\lambda s}}{2} \right) \right) \notag \\
&= \frac{d D_{\text{left}}}{d x_{\text{left}}} \bigg|_{x_\text{left}=x^{(1-\lambda)g+\lambda s}_\text{left}} = -\frac{d D_{\text{right}}}{d x_{\text{right}}} \bigg|_{x_\text{right}=x^{(1-\lambda)g+\lambda s}_\text{right}} \notag \\
&= -\frac{d D_{\text{right}}}{d x_{\text{right}}} \bigg|_{x_\text{left}=x^{g}_\text{right}} = \frac{d D_{\text{left}}}{d x_{\text{left}}} \bigg|_{x_\text{left}=x^{g}_\text{left}} \notag \\
&= \frac{1}{2} g \left( \frac{x_\text{left}^{g} + x_\text{right}^{g}}{2} \right) 
F_c\left( m\left( \frac{x_\text{left}^{g} + x_\text{right}^{g}}{2} \right) \right).
\label{eq:main}
\end{align}
For $\lambda>0$, this equation holds if and only if $\frac{1}{2} g \left( \frac{x_\text{left}^{g} + x_\text{right}^{g}}{2} \right) F_c\left( m\left( \frac{x_\text{left}^{g} + x_\text{right}^{g}}{2} \right) \right)=\frac{1}{2} s \left( \frac{x_\text{left}^{s} + x_\text{right}^{s}}{2} \right) 
F_c\left( m\left( \frac{x_\text{left}^{s} + x_\text{right}^{s}}{2} \right) \right)$. Under this condition, \(x_\text{left}^g\) and \(x_\text{right}^g\) would also be Nash equilibrium points of the distribution \(s\). However, this contradicts the uniqueness of the Nash equilibrium. Therefore, the initial assumption does not hold, and the proof is complete.

Furthermore, we can draw a more general conclusion: when a distribution \(g(x)\) converges to another distribution \(s(x)\), let the convergence sequence be denoted by \(\{g_i\}\). If any two distributions \(g_i\) and \(g_j\) (\(i < j\)) in this sequence can be expressed as \((1-\lambda_i) g + \lambda_i s\) and \((1-\lambda_j) g + \lambda_j s\), respectively, where \(\lambda_i < \lambda_j\), then, irrespective of the specific form of convergence, the Nash equilibrium points corresponding to \(g_j\), denoted as \(x_\text{left}^{g_j}\) and \(x_\text{right}^{g_j}\), must lie within the intervals:
$
[ \min(x_\text{left}^{g_i}, x_\text{left}^{h}), \max(x_\text{left}^{g_i}, x_\text{left}^{h}) ]
$
and 
$
[ \min(x_\text{right}^{g_i}, x_\text{right}^{h}), \max(x_\text{right}^{g_i}, x_\text{right}^{h}) ],
$
respectively.

To further investigate the influence of the voter distribution on the Nash equilibrium points, we employ a numerical example. Assume the voter distribution transitions from a single-peaked to a double-peaked shape: $g(x) = (1 - \alpha)\frac{1}{0.9256} \left(e^{-\frac{x^2}{2 \cdot 0.5^2}} - 0.1353\right) + \alpha\frac{1}{0.4669} \left(0.5 \cdot e^{-\frac{(x + 0.5)^2}{2 \cdot 0.3^2}} + 0.5 \cdot e^{-\frac{(x - 0.5)^2}{2 \cdot 0.3^2}}-0.1246 \right)$, where the parameter \(\alpha\) controls the transition from the single peak (\(\alpha = 0\)) to the double peak (\(\alpha = 1\)). The support of the distribution is \([-1, 1]\). The normalization ensures the total probability integrates to 1. Set \(F_c(x) = 0.5x\) and \(m(x) = x^2 + 0.5\). The deviation costs for the left-wing and right-wing parties are given by: $D_{\text{left}} = 0.3(x_{\text{left}} + 0.7)^2 \quad \text{and} \quad D_{\text{right}} = 0.5(x_{\text{right}} - 0.8)^2$.

We analyze the variation in Nash equilibrium points $x_{\text{left}}$ and $x_{\text{right}}$ across different configurations of \(\alpha\) (\( \alpha = 0, 0.5, 1\)), and examine the equilibrium dynamics through graphical illustrations. The figures depict:

\begin{figure}[H]
    \centering
    \begin{minipage}[b]{0.4\textwidth}
        \centering
        \includegraphics[width=\textwidth]{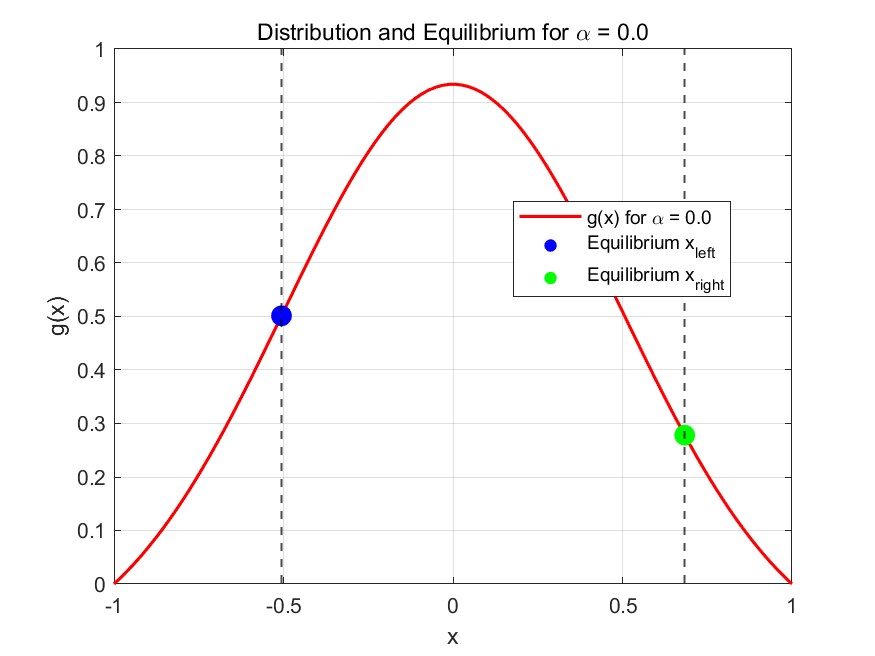}
        \caption{Distribution and Equilibrium for $\alpha = 0.0$}
        \label{fig:image1}
    \end{minipage}
    \hspace{0.05\textwidth}  
    \begin{minipage}[b]{0.4\textwidth}
        \centering
        \includegraphics[width=\textwidth]{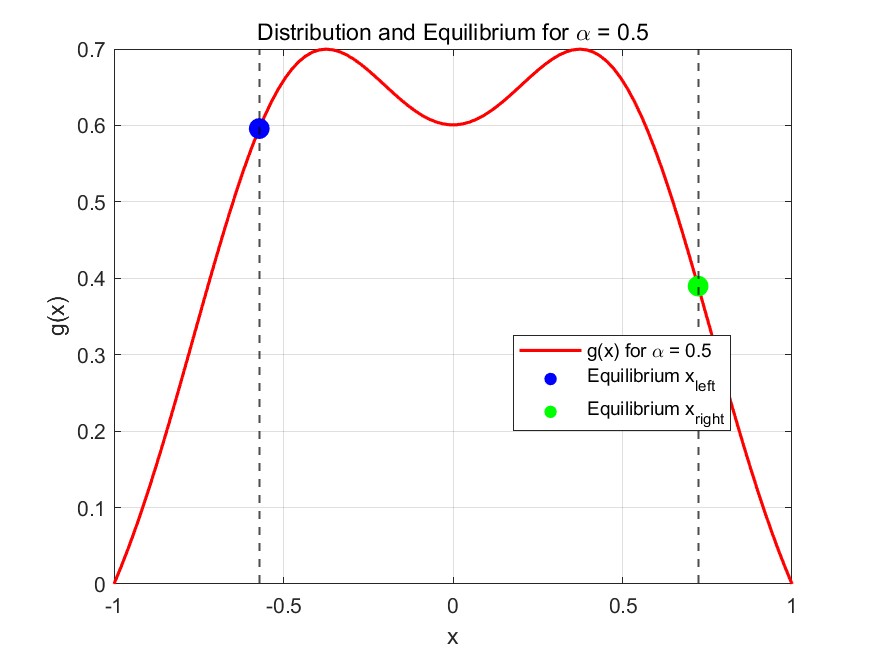}
        \caption{Distribution and Equilibrium for $\alpha = 0.5$}
        \label{fig:image2}
    \end{minipage}
 
\end{figure}

\begin{figure}[H]
    \centering
    \begin{minipage}[b]{0.4\textwidth}
        \centering
        \includegraphics[width=\textwidth]{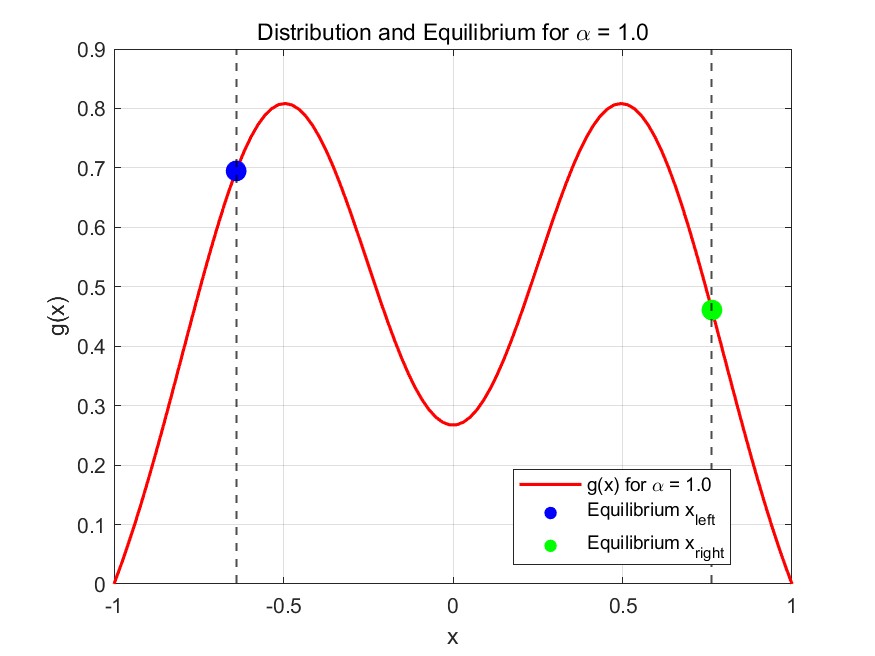}
        \caption{Distribution and Equilibrium for $\alpha = 1.0$}
        \label{fig:image3}
    \end{minipage}
    \hspace{0.05\textwidth}  
    \begin{minipage}[b]{0.4\textwidth}
        \centering
        \includegraphics[width=\textwidth]{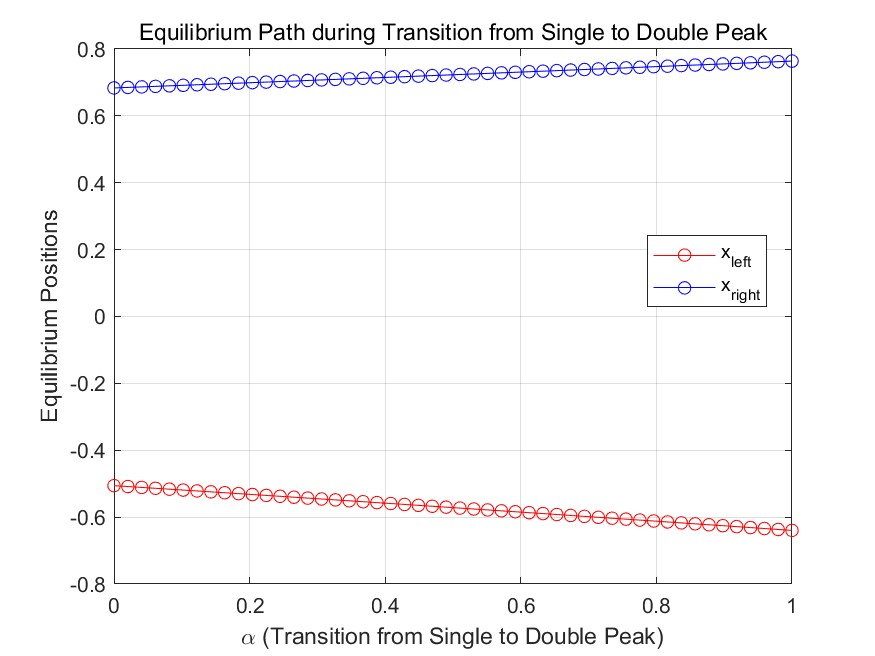}
        \caption{Equilibrium Path during Transition from Single to Double Peak}
        \label{fig:image4}
    \end{minipage}
 
\end{figure}

Generally speaking, when the voter distribution changes from a single-peaked distribution to a double-peaked distribution, the Nash equilibrium path tend to expand to two poles due to decrease of $g\left( \frac{x_{\text{left}} + x_{\text{right}}}{2} \right) F_c\left( m\left( \frac{x_{\text{left}} + x_{\text{right}}}{2} \right) \right)$. Intuitively, as the distribution of voters changes from a single peak to a double peak, the influence of the middle voters gradually decreases. Both parties no longer need to get closer to the middle voters, but to get closer to their own ideal points to consolidate their base. But this does not necessarily happen, specific situation still depends on the quantitative analysis above. The following is an example of a single-peak distribution transforminging into a double-peak distribution while equilibrium points moving towards the y-axis:  $g(x) = (1 - \alpha)\frac{1}{0.9235} \left(e^{-\frac{x^2}{16}} - 0.9394\right) + \alpha \frac{1}{0.4682} \left(0.5 \cdot e^{-\frac{(x + 0.3)^2}{2 \cdot 0.25^2}} + 0.5 \cdot e^{-\frac{(x - 0.3)^2}{2 \cdot 0.25^2}} - 0.0099\right)$, 
where the parameter $\alpha$ controls the transition from the single peak ($\alpha = 0$) to the double peak ($\alpha = 1$). Set $F_c(x) = \min(0.5 \cdot x, 1)$ and $m(x) = x^2 + 0.7$, respectively. The deviation costs for the left-wing and right-wing parties are given by 
$D_{\text{left}} = 0.4(x_{\text{left}} + 0.7)^2$ and $D_{\text{right}} = 0.5(x_{\text{right}} - 0.6)^2$.

\begin{figure}[H]
    \centering
    \begin{minipage}[b]{0.4\textwidth}
        \centering
        \includegraphics[width=\textwidth]{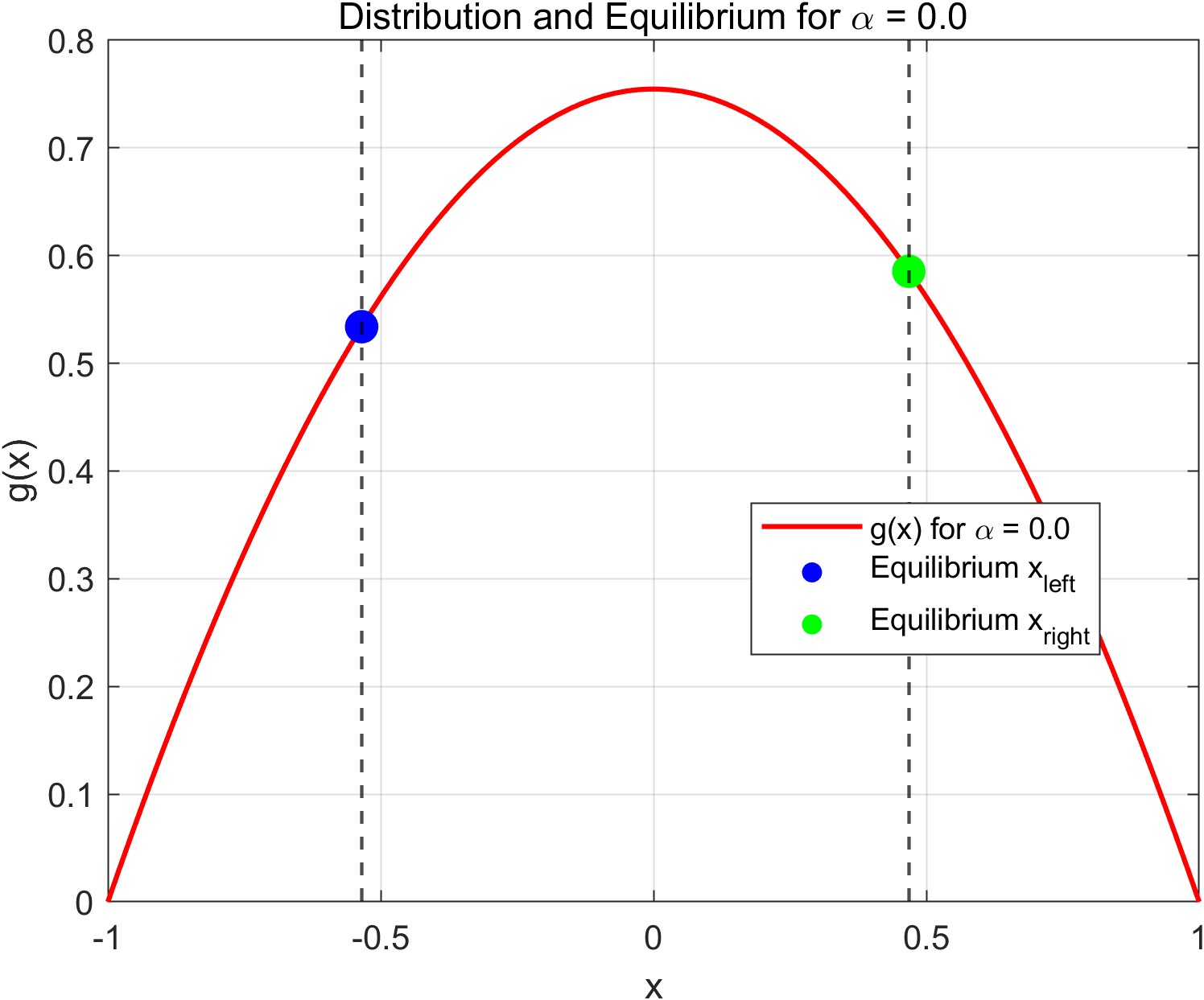}
        \caption{Distribution and Equilibrium for $\alpha = 0.0$}
        \label{fig:image1}
    \end{minipage}
    \hspace{0.05\textwidth}  
    \begin{minipage}[b]{0.4\textwidth}
        \centering
        \includegraphics[width=\textwidth]{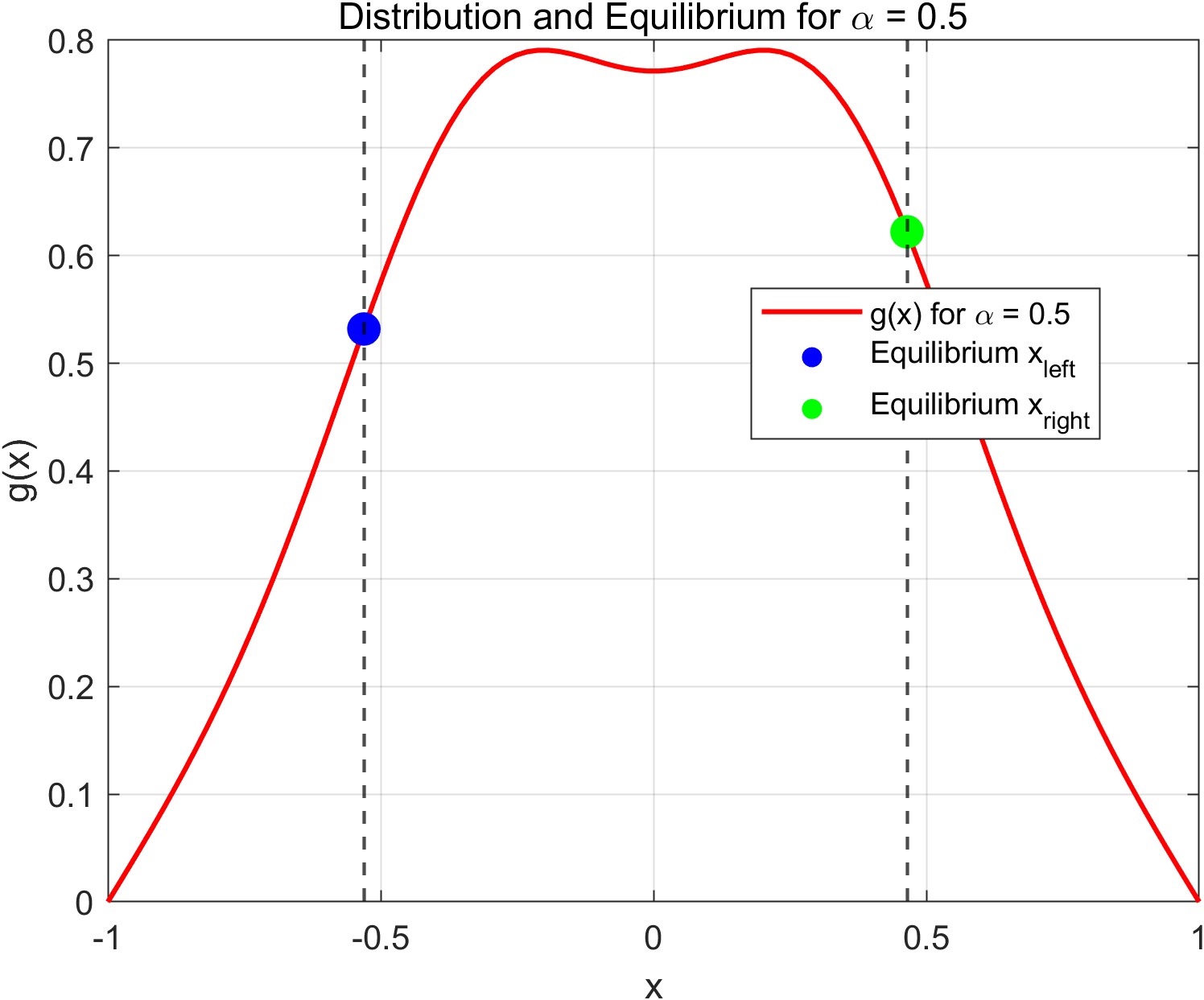}
        \caption{Distribution and Equilibrium for $\alpha = 0.5$}
        \label{fig:image2}
    \end{minipage}
 
\end{figure}

\begin{figure}[H]
    \centering
    \begin{minipage}[b]{0.4\textwidth}
        \centering
        \includegraphics[width=\textwidth]{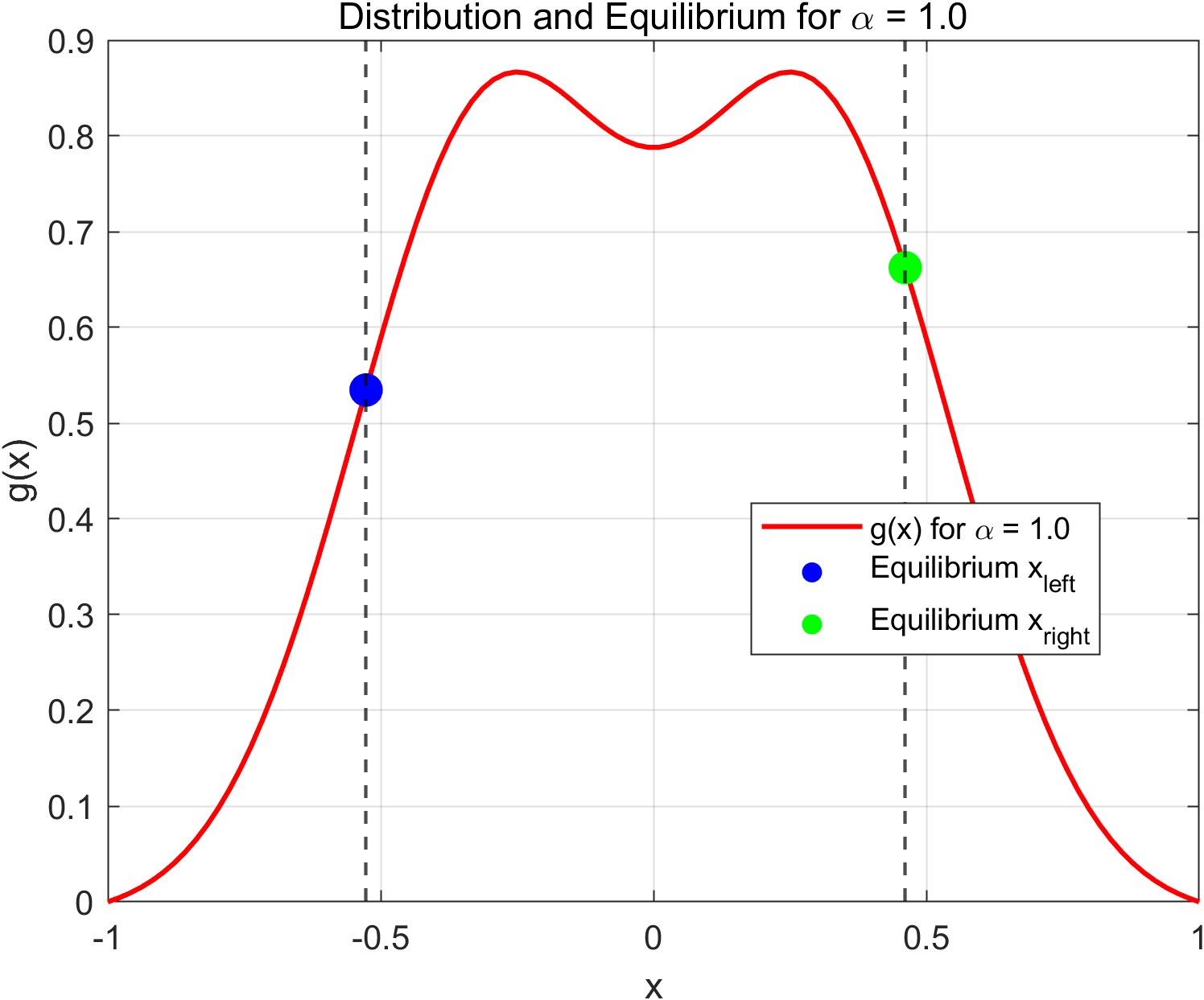}
        \caption{Distribution and Equilibrium for $\alpha = 1.0$}
        \label{fig:image3}
    \end{minipage}
    \hspace{0.05\textwidth}  
    \begin{minipage}[b]{0.4\textwidth}
        \centering
        \includegraphics[width=\textwidth]{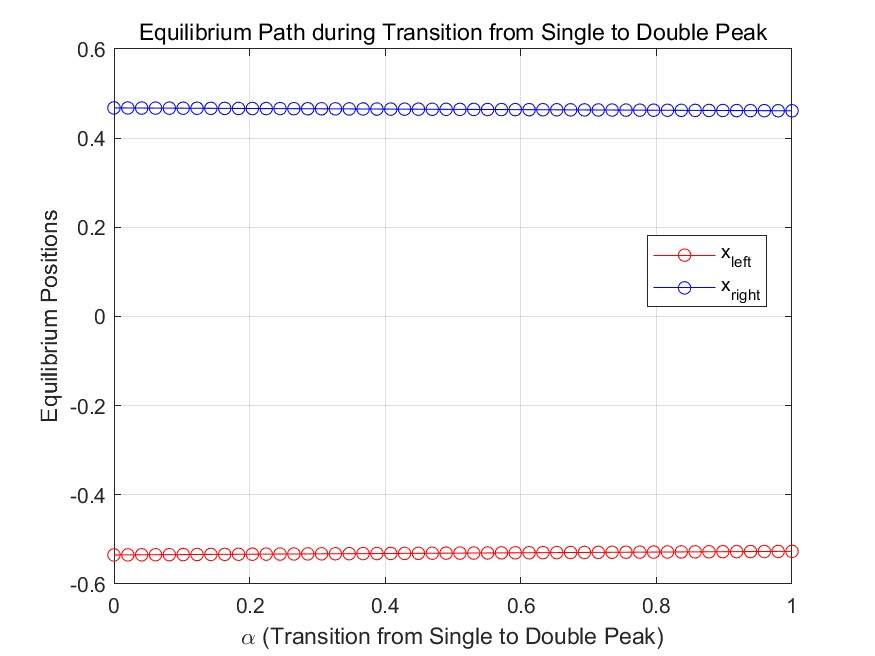}
        \caption{Equilibrium Path during Transition from Single to Double Peak}
        \label{fig:image4}
    \end{minipage}
 
\end{figure}

\subsection{True Voter Distribution Unknown}
In practice, it is nearly impossible to obtain the exact distribution of voter preferences. However, we can strive to approximate it as accurately as possible. To begin, consider the decision-making process of one party given the exogenous choice of the other. Let the true voter distribution be denoted as \(G\), with its probability density function (PDF) given by \(g(x)\). Due to the noise inherent in information transmission, such as that introduced by polling, both parties receive noisy approximations of the true distribution, denoted as \(P_\text{left}\) and \(P_\text{right}\), with their respective PDFs \(p_\text{left}(x)\) and \(p_\text{right}(x)\). These approximated distributions satisfy the following conditions:

\begin{equation}
D(g(x), p_\text{left}(x)) < \alpha_\text{left}, \quad D(g(x), p_\text{right}(x)) < \alpha_\text{right},
\end{equation}
where $D$ is a functional that measures the difference between two distributions, and $\alpha_\text{left}$ and $\alpha_\text{right}$ are public information. Common metrics for $D$ include Kullback-Leibler (KL) Divergence, Wasserstein Distance, and Jensen-Shannon (JS) Divergence. Let $\hat{\mathcal{G}}_{\alpha, p}$ denote the set of distributions $\hat{g}(x)$ satisfying $D(\hat{g}, p) < \alpha$. We assume that every element of $\hat{\mathcal{G}}_{\alpha, p}$ meets the conditions for the existence and uniqueness of a Nash equilibrium.

Next, we attempt to construct a \textit{Bayesian Nash Equilibrium} (BNE) for the two parties. Given the noisy distributions $p_\text{left}$ and $p_\text{right}$, the BNE is defined as follows:

\begin{equation}
x_\text{left}^*(p_\text{left}) = \arg\max_{x_\text{left}} \mathbb{E}_{\hat{g}, x_\text{right}^*} \left[ U_\text{left}(x_\text{left}, x_\text{right}^*, \hat{g}) \right],
\end{equation}
\begin{equation}
x_\text{right}^*(p_\text{right}) = \arg\max_{x_\text{right}} \mathbb{E}_{\hat{g}, x_\text{left}^*} \left[ U_\text{right}(x_\text{left}^*,x_\text{right}, \hat{g}) \right].
\end{equation}

To derive an analytical expression for the BNE, consider $x_\text{left}^*(p_\text{left})$ as an example. It can be expressed as: 
\begin{equation}
x_\text{left}^*(p_\text{left}) = \arg\max_{x_\text{left}} \int_{\hat{g}} \int_{p_\text{right}} U_\text{left}(x_\text{left}, x_\text{right}^*(p_\text{right}), \hat{g}) P(p_\text{right} \mid \hat{g}) P(\hat{g} \mid p_\text{left}) \,dp_\text{right} d\hat{g}.
\end{equation}
Here, $P(\hat{g} \mid p_\text{left})$ is the PDF of $\hat{g} \in \hat{\mathcal{G}}_{\alpha_\text{left}, p_\text{left}}$, and $P(p_{\text{right}} \mid \hat{g})$ is the PDF of the right party's optimal strategy $p_\text{right}$, where $p_\text{right} \in \hat{\mathcal{G}}^{-1}_{\alpha_\text{right}, \hat{g}}$. The set $\hat{\mathcal{G}}^{-1}_{\alpha_\text{right}, \hat{g}}$ contains all distributions $p_\text{right}$ such that $\hat{g} \in \hat{\mathcal{G}}_{\alpha_\text{right}, p_\text{right}}$. The conditional PDF $P(p_\text{right} \mid \hat{g})$ can be derived via Bayes' theorem, allowing us to compute $P(x_{\text{right}}^* \mid \hat{g})$.
Similarly, $x_\text{right}^*(p_\text{right})$ can be witten as:
\begin{equation}
x_\text{right}^*(p_\text{right})=\arg\max_{x_{\text{right}}} 
\int_{\hat{g}} \int_{p_{\text{left}}} 
U_\text{right}(x_{\text{left}}^*(p_\text{left}), x_{\text{right}}, \hat{g}) 
P(p_{\text{left}} \mid \hat{g}) 
P(\hat{g} \mid p_{\text{right}}) 
\, dp_{\text{left}} d\hat{g}.
\end{equation}

This formalism ensures a comprehensive framework for deriving the BNE in the presence of noisy information transmission. Therefore, even when the true distribution is unknown and there is no prior information of distribution, the two parties can still reach a BNE. 

Here, we do not adopt the concept of information cost commonly used in information economics for the following reasons:  
\begin{enumerate}
    \item In real-world polling, political parties strive to obtain the voter distribution as accurately as possible, irrespective of cost.  
    \item The accuracy of the obtained voter distribution is not necessarily correlated with the associated information cost.  
    \item Even if information cost is positively correlated with the accuracy of the voter distribution, such costs are typically represented by financial and human resources invested in conducting polls, which are largely irrelevant to the utility framework within the model.
\end{enumerate}

\section{Mutiple Dimensions Model}
We are transitioning from a one-dimensional analysis of economic ideology to a multi-dimensional framework. In this expanded model, we introduce a new factor: the feasibility of economic ideologies across different dimensions.

For instance, traditional right-wing economic ideologies tend to advocate for low taxes and low welfare, while traditional left-wing ideologies favor high taxes and high welfare. Although debates between these two schools of thought persist, both sides generally acknowledge that the other's policies are practically feasible.

In contrast, the feasibility of policies advocating for high welfare alongside low taxes is highly questionable. As the well-known proverb goes, ``There’s no such thing as a free lunch." Historically, such a combination often results in fiscal collapse or rampant inflation—frequently both at once. Therefore, we write the utility function of the previous one-dimensional economic ideology into a multidimensional form represented by a vector:

\begin{equation}
U_{\text{left}}(\mathbf{x}_{\text{left}}, \mathbf{x}_{\text{right}}) = 
\int_{\mathcal{Z}_{\text{left}}} g(\mathbf{x}) F_c(m(\mathbf{x})) \, d\mathbf{x} 
- D_{\text{left}}(\mathbf{x}_{\text{left}}, \mathbf{ideal}_{\text{left}}) 
- \Phi(\mathbf{x}_{\text{left}}),
\end{equation}
\begin{equation}
U_{\text{right}}(\mathbf{x}_{\text{left}}, \mathbf{x}_{\text{right}}) = 
\int_{\mathcal{Z}_{\text{right}}} g(\mathbf{x}) F_c(m(\mathbf{x})) \, d\mathbf{x} 
- D_{\text{right}}(\mathbf{x}_{\text{right}}, \mathbf{ideal}_{\text{right}}) 
- \Phi(\mathbf{x}_{\text{right}}),
\end{equation}
 where \( \mathcal{Z}_{\text{left}} \) and \( \mathcal{Z}_{\text{right}} \) are the integration domains for the left-wing and right-wing ideologies, respectively. For example, \( \mathcal{Z}_{\text{left}} = \{ \mathbf{x} \in \mathbb{R}^n \mid \mathbf{x} \in [-1, (\mathbf{x}_{\text{left}} + \mathbf{x}_{\text{right}})/2] \} \) and \( \mathcal{Z}_{\text{right}} = \{ \mathbf{x} \in \mathbb{R}^n \mid \mathbf{x} \in [ (\mathbf{x}_{\text{left}} + \mathbf{x}_{\text{right}})/2,1] \} \); \( \Phi \) is the feasibility indicator for the internal coherence of the ideology \( \mathbf{x}_{\text{left}} \) or \( \mathbf{x}_{\text{right}} \), representing the overall practicality and viability of implementing the proposed policy mix. The higher the feasibility of \(\mathbf{x}\), the smaller \( \Phi \). Like the possible indifference feasibility curve below, we suppose that the horizontal axis from left to right indicates that the tax level is getting higher and higher; the vertical axis from top to bottom indicates that the welfare level is getting higher and higher. The process of the curve changing from yellow to blue indicates that feasibility gradually increases, that is, from the extreme of low taxes and high welfare (1,-1) to the extreme of high taxes and low welfare (-1,1).  
\begin{figure}[H]
    \centering
    \includegraphics[width=0.5\textwidth]{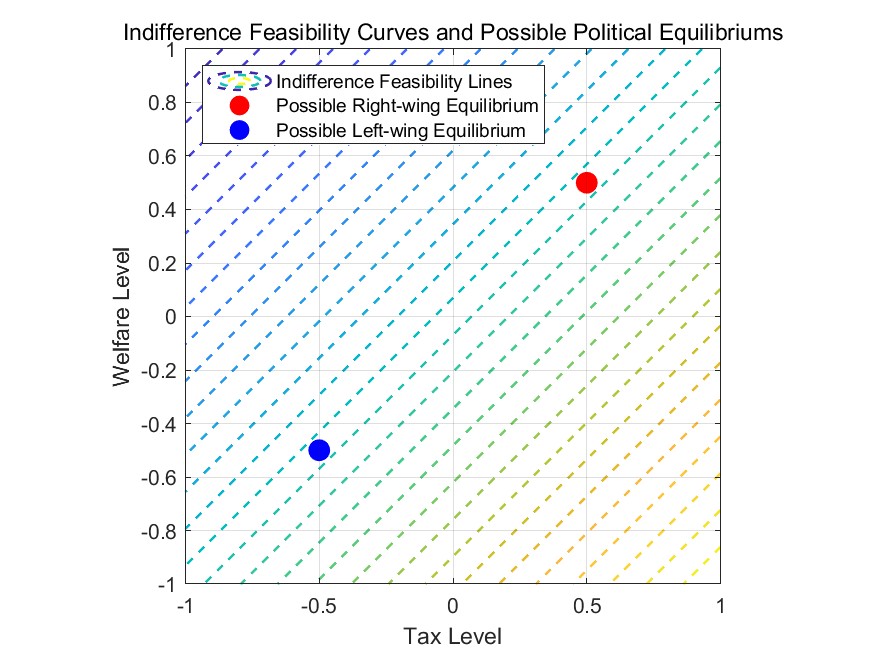}
    \caption{Indifference Feasibility Curves and Possible Political Equilibriums}
    \label{fig:image1}
\end{figure}
Here, $\nabla_{\text{tax}} \Phi > 0, \quad \nabla_{\text{welfare}} \Phi < 0$, which means that reducing financial resources will increase feasibility costs, while reducing fiscal expenditures will reduce feasibility costs.

Assuming that the utility function of such a multidimensional economic ideology also satisfies the existence and uniqueness conditions of Nash equilibrium: Non-empty, Compact, and Convex Strategy Spaces, Continuity and Quais-concavity of Utility Function, the Nash equilibrium satisfies:

\begin{equation}
\nabla_{\mathbf{x}_{\text{left}}} U_{\text{left}}(\mathbf{x}_{\text{left}}^*, \mathbf{x}_{\text{right}}^*) = \mathbf{0}, \quad
\nabla_{\mathbf{x}_{\text{right}}} U_{\text{right}}(\mathbf{x}_{\text{left}}^*, \mathbf{x}_{\text{right}}^*) = \mathbf{0}.
\end{equation}
Intuitively, the Nash equilibrium between the two parties is to achieve a balance between voters, ideal ideology and actual policy feasibility.

Now we take a two-demensional strategy for utility function as an example. Suppose that
$
G(x_1, x_2) = 
\frac{\exp\left(-\frac{(x_1 - \mu_1)^2}{2\sigma_1^2}\right)}{\int_{-1}^{1} \exp\left(-\frac{(x - \mu_1)^2}{2\sigma_1^2}\right) \, dx}
\cdot
\frac{\exp\left(-\frac{(x_2 - \mu_2)^2}{2\sigma_2^2}\right)}{\int_{-1}^{1} \exp\left(-\frac{(x - \mu_2)^2}{2\sigma_2^2}\right) \, dx}$, $\mu_1 = 0.1$, $\sigma_1 = 0.5$, $\mu_2 = -0.1,$ $\sigma_2 = 0.5$, $F_c(x) = x$, $m(x_1,x_2) = 0.25(x_1^2 +x_2^2+2)$, $D_{\text{left}}=(x_1+0.7)^2+(x_2+0.5)^2$,$D_{\text{right}}=(x_1-0.6)^2+(x_2-0.6)^2$, $\Phi=0.1(x_1-x_2)$. Then we get a pair of Nash equilibrium points in the joint voter distribution:

\begin{figure}[H]
    \centering
    \includegraphics[width=0.5\textwidth]{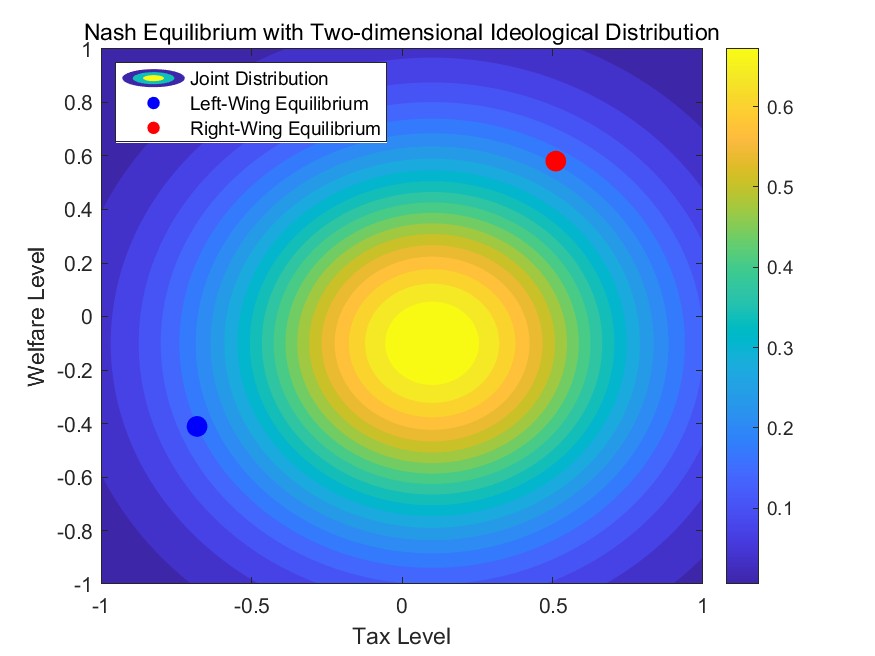}
    \caption{Nash Equilibrium with Two-dimensional Ideological Distribution}
    \label{fig:image1}
\end{figure}

Now we introduce a small perturbation \(\alpha\) of feasibility cost \(\Phi\) then analyze the elasticity of equilibrium through Taylor expansion:

\begin{equation}
H_{\text{left-left}} \Delta \mathbf{x}_{\text{left}} + C_{\text{left-right}} \Delta \mathbf{x}_{\text{right}} = \alpha \nabla_{\mathbf{x}_{\text{left}}} \Phi(\mathbf{x}_{\text{left}}^*), 
\end{equation}
\begin{equation}
C_{\text{right-left}} \Delta \mathbf{x}_{\text{left}} + H_{\text{right-right}} \Delta \mathbf{x}_{\text{right}} = \alpha \nabla_{\mathbf{x}_{\text{right}}} \Phi(\mathbf{x}_{\text{right}}^*). 
\end{equation}
where: $H_{\text{left-left}} = \nabla^2_{\mathbf{x}_{\text{left}}} U_{\text{left}}$, $H_{\text{right-right}} = \nabla^2_{\mathbf{x}_{\text{right}}} U_{\text{right}}$, $C_{\text{left-right}} = \nabla^2_{\mathbf{x}_{\text{left}} \mathbf{x}_{\text{right}}} U_{\text{left}}$, $C_{\text{right-left}} = \nabla^2_{\mathbf{x}_{\text{right}} \mathbf{x}_{\text{left}}} U_{\text{right}}$.

In matrix form, the system can be written as:
\begin{equation}
\mathbf{H} \cdot \Delta \mathbf{x} = \alpha \mathbf{b},
\end{equation}
where:
\[
\mathbf{H} =
\begin{bmatrix}
H_{\text{left-left}} & C_{\text{left-right}} \\
C_{\text{right-left}} & H_{\text{right-right}}
\end{bmatrix}, \quad
\Delta \mathbf{x} =
\begin{bmatrix}
\Delta \mathbf{x}_{\text{left}} \\
\Delta \mathbf{x}_{\text{right}}
\end{bmatrix}, \quad
\mathbf{b} =
\begin{bmatrix}
\nabla_{\mathbf{x}_{\text{left}}} \Phi(\mathbf{x}_{\text{left}}^*) \\
\nabla_{\mathbf{x}_{\text{right}}} \Phi(\mathbf{x}_{\text{right}}^*)
\end{bmatrix}.
\]
Then, the equilibrium variation is given by:
\begin{equation}
\Delta \mathbf{x} = \alpha \mathbf{H}^{-1} \cdot \mathbf{b},
\end{equation}
while equilibrium elasticity is
\begin{equation}
	E_\mathbf{x} = \mathbf{H}^{-1} \cdot \mathbf{b} \cdot\mathbf{x}'^{-1},
\end{equation}
where $\mathbf{x}=
\begin{bmatrix}
\mathbf{x}^*_{\text{left}} \\
\mathbf{x}^*_{\text{right}}
\end{bmatrix}. $

For the convenience of analysis, we may ignore the cross-influence terms \(C_{\text{left-right}}\) and \(C_{\text{right-left}}\), as their impact is smaller compared to the semi-negative definite matrices \(H_{\text{left-left}}\) and \(H_{\text{right-right}}\). Under this assumption, the changes in the decision variables can be expressed as $\Delta \mathbf{x}_{\text{left}} = \alpha H_{\text{left-left}}^{-1} \nabla_{\mathbf{x}_{\text{left}}} \Phi(\mathbf{x}_{\text{left}}^*), \quad
\Delta \mathbf{x}_{\text{right}} = \alpha H_{\text{right-right}}^{-1} \nabla_{\mathbf{x}_{\text{right}}} \Phi(\mathbf{x}_{\text{right}}^*).$

In most cases, we can further simplify the analysis by ignoring the cross-effects of other dimensions in the vector: $\Delta \mathbf{x}_{\text{left},i} = \alpha \frac{\partial^2 x_{\text{left},i}}{\partial^2 U_\text{left}} \nabla_{x_{\text{left},i}} \Phi$, $\Delta \mathbf{x}_{\text{right},i} = \alpha \frac{\partial^2 x_{\text{right},i}}{\partial^2 U_\text{right}} \nabla_{x_{\text{right},i}} \Phi.$ This assumption is valid when the off-diagonal elements of the matrices \(H_{\text{left-left}}\) and \(H_{\text{right-right}}\), as well as the cross-influence terms \(C_{\text{left-right}}\) and \(C_{\text{right-left}}\), are relatively small compared to the diagonal elements. Such cases typically arise when the interaction between different variables or dimensions is weak, allowing the effects of each variable to be approximated independently. For example, this simplification is reasonable when variables are nearly orthogonal or the magnitude of their coupling is negligible in the context of the model's overall dynamics.

Next, we still use the example above but put a perturbation $\alpha$ on $\Phi$ from 0.5 to 2. Then we calculate the equilibrium path below.
 
\begin{figure}[H]
    \centering
    \includegraphics[width=0.5\textwidth]{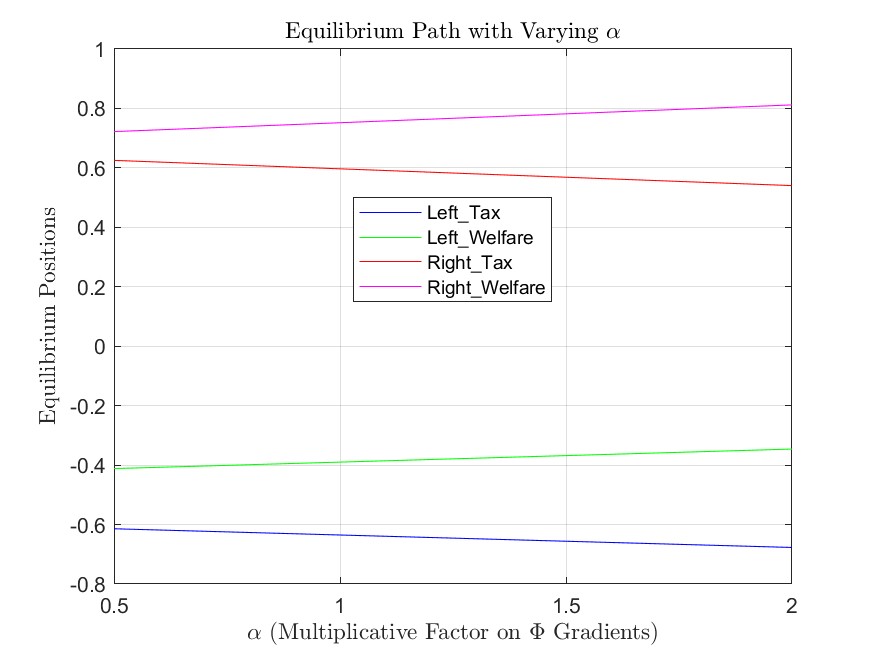}
    \caption{Equilibrium Path with Varying $\alpha$}
    \label{fig:image1}
\end{figure}

As shown in the figure, when the feasibility cost changes while all other conditions remain constant, the equilibrium path's direction of change aligns with the case where all cross-effects are ignored. Specifically, when the feasibility cost increases, both the left and right parties are motivated to reduce welfare and increase taxes. Conversely, when the feasibility cost decreases, both parties are inclined to increase welfare and lower taxes.

This explains the widespread fiscal crisis in today's democratic countries: due to the development of the globalization, governments in democratic countries have more powerful means to expand fiscal space, such as issuing bonds and loans, which has led to an overall reduction in feasibility costs. Taking the level of taxation and welfare as an example, both left-wing and right-wing parties tend advocate more welfare or less taxation. At least, it is easy for a left-wing government to talk about increasing welfare, but difficult to implement when it comes to increasing taxes; similarly, it is easy for a right-wing government to talk about reducing taxes, but hesitant when it comes to reducing welfare. Therefore, regardless of whether the left-wing party or right-wing party is in power, there is pressure to further increase the national fiscal deficit.   

Greece serves as a compelling case study on this issue. During the Greek debt crisis, the right-wing government implemented austerity measures, advocating for welfare cuts to address the fiscal imbalance. This sparked widespread public protests, ultimately leading to the overthrow of the right-wing administration. Subsequently, a government led by Alexis Tsipras, representing the radical left, came to power. On the surface, Tsipras's anti-austerity stance and commitment to preserving welfare align with leftist principles. However, his administration notably refrained from advocating for higher taxes, a traditional leftist economic policy. The rationale is clear: championing higher taxes would have undermined Tsipras's electoral appeal, as the public's mandate was to maintain welfare, not to increase their tax burden. Faced with this dilemma, Tsipras’s government sought an alternative solution—defaulting on debt obligations to the European Union. 

Before Greece joined the Eurozone, similar situations had occurred but were not as severe as during the Greek crisis. This was because Greece’s economic system, operating under the drachma, was relatively independent, and the policy feasibility cost was higher. When deficits grew excessive, the drachma would depreciate, effectively imposing economic discipline across the country. This mechanism allowed the Greek government to maintain its finances at a relatively sustainable level. However, with the adoption of the euro, which would not depreciate painfully, Greece found itself slowly slipping into the trap of crisis in the sweet dreamland of high welfare and low taxes.
 
\section{Incompatibility with Community Ideology}
After analyzing the model of economic ideology, we now turn to a subject that traditional voting theories often overlook or conflate with economic ideology: community ideology. Below, we summarize the key differences between economic ideology and community ideology as they manifest in political life.

First, economic differences are relatively easy to quantify, whereas differences between communities are more ambiguous. For instance, if a left-wing party advocates for a 30\% tax rate and a right-wing party advocates for a 20\% tax rate, voters can more easily decide which policy aligns with their preferences (or at least bothers them less). However, when opposing ethnic groups or rival religions accuse each other of crimes, these conflicts are much harder to measure or resolve. 

More importantly, voters' economic ideologies are fluid in the real world. For example, someone who moves from a lower income bracket to the middle class may shift from desiring greater public benefits to favoring lower taxes. On a societal level, if laissez-faire policies exacerbate inequality, public sentiment may swing toward economic policies that emphasize fairness. In contrast, transformation within a community is almost impossible. A Muslim is unlikely to convert to Christianity; an Irishman cannot become British. Even though gender transitions have become more common in recent years, such transformations remain far from universal. This fixed identity within communities often makes it harder for individuals to approach issues objectively. Historically, conflicts between communities—whether based on ethnicity, race, or religion—have tended to be longer-lasting, larger in scale, and more violent than class-based conflicts. Consequently, voters are more likely to abstain from voting when both parties' positions deviate too far from their community's ideology. For instance, extremists within a community often regard moderates who seek reconciliation with a hostile community as “traitors” and may even despise them more than their adversaries. Here, the mechanism of homophily, or the tendency for individuals to associate with similar others, as proposed by Fu et al. (2012) and McPherson et al. (2001), is particularly evident. Within communities, the diffusion of associated beliefs and the mechanisms proposed by DellaPosta (2020), DellaPosta et al. (2015), and Goldberg and Stein (2018) further exacerbate ideological differences between groups, intensifying polarization. In summary, differences in community ideology are significantly more exclusive and irreconcilable than those in economic ideology.

Another important feature of political parties' behavior regarding community ideology is that, within the framework of modern democracy, a party rarely adopts the ideology of a specific community as its inherent or official value. Even when a political party's support base overlaps significantly with a particular community, it generally avoids identifying itself solely with that group. For example, the African National Congress in South Africa is primarily supported by black nationalists but does not identify as a “black political party.” Similarly, while the Bharatiya Janata Party and the Indian National Congress each have distinct voting bases rooted in specific communities, both claim to represent the interests of all Indian citizens. Therefore, it is difficult for us to identify a political party as representing a specific community ideology.  

Lastly, as Benedict Anderson proposed, a nation is an ``imagined community." The formation of community ideology heavily depends on the interaction between leaders and their supporters, as this interaction underpins the collective imagination. While the influence of political parties on voters' economic ideology can often be overlooked in certain analyses, the construction and dissemination of community ideology are largely driven by political leaders, who actively shape and instill various imagined ideas within groups.

For these reasons, the model we developed to describe party choices in economic ideology cannot be directly applied to the analysis of community ideology. Therefore, it is necessary to establish a new framework of thought.

\section{Conclusion}
In this paper, we revisited the limitations of the Median Voter Theorem and proposed a novel framework to analyze the optimal ideological positions of political parties. By incorporating Nash equilibrium into the model, we explored the mechanisms through which deviation costs, voter distribution, and policy feasibility shape parties' strategic choices. Additionally, we demonstrated that even in cases where the voter distribution is unknown, optimal ideological positions for both parties can still be identified. Moreover, we highlighted why our model cannot be applied to community ideologies, emphasizing the fundamental differences between economic and community ideological frameworks.

Our research contributes to a deeper understanding of the forces driving the ideological positioning of political parties and provides a theoretical foundation for future empirical studies. By connecting theoretical insights with practical implications, we aim to offer valuable perspectives for political scientists and policymakers navigating the complexities of modern political systems.

\section*{References}
\begin{enumerate}
    \item Hotelling, H. (2024). Stability in Competition1. In \textit{The Foundations of Price Theory Vol 4} (pp. 241-260). Routledge.
    \item Downs, A. (1957). An Economic Theory of Political Action in a Democracy. \textit{Journal of Political Economy, 65}(2), 135-150.
    \item Palfrey, T. R. (1984). Spatial equilibrium with entry. The Review of Economic Studies, 51(1), 139-156.
    \item Enelow, J. M., \& Hinich, M. J. (1984). \textit{The spatial theory of voting: An introduction}. CUP Archive.
    \item Gomez, B. T., Hansford, T. G., \& Krause, G. A. (2007). The Republicans should pray for rain: Weather, turnout, and voting in US presidential elections. \textit{The Journal of Politics, 69}(3), 649-663.
    \item Fenster, M. J. (1994). The impact of allowing day of registration voting on turnout in US elections from 1960 to 1992: A research note. \textit{American Politics Quarterly, 22}(1), 74-87.
    \item Davis, O. A., \& Hinich, M. J. (1965). A Mathematical Model of Policy Formation in a Democratic Society. \textit{Graduate School of Industrial Administration, Carnegie Institute of Technology}.
    \item Hinich, M. J., \& Ordeshook, P. C. (1970). Plurality Maximization vs. Vote Maximization: A Spatial Analysis with Variable Participation. \textit{American Political Science Review, 64}(3), 772-791.
    \item Coughlin, P. J. (1992). \textit{Probabilistic Voting Theory}. Cambridge University Press.
    \item Burden, B. C. (1997). Deterministic and Probabilistic Voting Models. \textit{American Journal of Political Science}, 1150-1169.
    \item Banks, J. S., \& Duggan, J. (2005). Probabilistic Voting in the Spatial Model of Elections: The Theory of Office-Motivated Candidates. In \textit{Social Choice and Strategic Decisions: Essays in Honor of Jeffrey S. Banks} (pp. 15-56). Berlin, Heidelberg: Springer Berlin Heidelberg.
    \item McKelvey, R. D., \& Patty, J. W. (2006). A Theory of Voting in Large Elections. \textit{Games and Economic Behavior, 57}(1), 155-180.
    \item Jones, M. I., Sirianni, A. D., \& Fu, F. (2022). Polarization, Abstention, and the Median Voter Theorem. \textit{Humanities and Social Sciences Communications, 9}(1), 1-12.
    \item Kollman, K., Miller, J. H., \& Page, S. E. (1992). Adaptive Parties in Spatial Elections. \textit{American Political Science Review, 86}(4), 929-937.
    \item Druckman, J. N., Peterson, E., \& Slothuus, R. (2013). How Elite Partisan Polarization Affects Public Opinion Formation. \textit{American Political Science Review, 107}(1), 57-79.
    \item Banda, K. K., \& Cluverius, J. (2018). Elite Polarization, Party Extremity, and Affective Polarization. \textit{Electoral Studies, 56}, 90-101.
    \item Rogowski, J. C., \& Langella, S. (2015). Primary Systems and Candidate Ideology: Evidence from Federal and State Legislative Elections. \textit{American Politics Research, 43}(5), 846-871.
    \item Mullinix, K. J. (2016). Partisanship and Preference Formation: Competing Motivations, Elite Polarization, and Issue Importance. \textit{Political Behavior, 38}, 383-411.
    \item Webster, S. W., \& Abramowitz, A. I. (2017). The Ideological Foundations of Affective Polarization in the U.S. Electorate. \textit{American Politics Research, 45}(4), 621-647.
    \item Black, D. (1958). \textit{The Theory of Committees and Elections}. Cambridge University Press.
    \item Fiorina, M. P., \& Abrams, S. J. (2008). Political Polarization in the American Public. \textit{Annual Review of Political Science, 11}(1), 563-588.
    \item Baldassarri, D., \& Bearman, P. (2007). Dynamics of Political Polarization. \textit{American Sociological Review, 72}(5), 784-811.
    \item Dixit, A. K., \& Weibull, J. W. (2007). Political Polarization. \textit{Proceedings of the National Academy of Sciences, 104}(18), 7351-7356.
    \item Grosser, J., \& Palfrey, T. R. (2014). Candidate Entry and Political Polarization: An Antimedian Voter Theorem. \textit{American Journal of Political Science, 58}(1), 127-143.
    \item Conover, M., Ratkiewicz, J., Francisco, M., Gonçalves, B., Menczer, F., \& Flammini, A. (2011). Political Polarization on Twitter. In \textit{Proceedings of the International AAAI Conference on Web and Social Media} (Vol. 5, No. 1, pp. 89-96).
    \item Bail, C. A., Argyle, L. P., Brown, T. W., Bumpus, J. P., Chen, H., Hunzaker, M. F., ... \& Volfovsky, A. (2018). Exposure to Opposing Views on Social Media Can Increase Political Polarization. \textit{Proceedings of the National Academy of Sciences, 115}(37), 9216-9221.
    \item Wang, X., Sirianni, A. D., Tang, S., Zheng, Z., \& Fu, F. (2020). Public discourse and social network echo chambers driven by socio-cognitive biases. \textit{Physical Review X, 10}(4), 041042.
    \item Fu, F., Nowak, M. A., Christakis, N. A., \& Fowler, J. H. (2012). The evolution of homophily. Sci. Rep. 2, 845.
    \item McPherson, M., Smith-Lovin, L., \& Cook, J. M. (2001). Birds of a feather: Homophily in social networks. Annual review of sociology, 27(1), 415-444.
    \item McPherson, M., Smith-Lovin, L., \& Cook, J. M. (2001). Birds of a feather: Homophily in social networks. Annual review of sociology, 27(1), 415-444.
    \item DellaPosta, D., Shi, Y., \& Macy, M. (2015). Why do liberals drink lattes?. American Journal of Sociology, 120(5), 1473-1511.
    \item Goldberg, A., \& Stein, S. K. (2018). Beyond social contagion: Associative diffusion and the emergence of cultural variation. American Sociological Review, 83(5), 897-932.
    \item Anderson, B. (2020). Imagined communities: Reflections on the origin and spread of nationalism. In The new social theory reader (pp. 282-288). Routledge.

\end{enumerate}

\end{document}